\documentclass{article}
\usepackage{epsfig}
\usepackage{cite}
\def\lapprox{\lower .7ex\hbox{$\;\stackrel{\textstyle <}{\sim}\;$}}
\def\gapprox{\lower .7ex\hbox{$\;\stackrel{\textstyle >}{\sim}\;$}}

\def\e{\epsilon}
\def\d{{\rm d}}

\def\sab{s_{12}}
\def\sac{s_{13}}
\def\sbc{s_{23}}

\def\sabc{s_{123}}

\newcommand{\trianglecrossNLO}[3]{
\mbox{\parbox{3cm}{\hspace{0.25cm}
\begin{picture}(2.5,1.4)
\thicklines
\put(0.3,0.7){\vector(1,0){0.1}}
\put(1.9,0.2){\vector(1,0){0.1}}
\put(1.9,1.2){\vector(1,0){0.1}}
\put(0,0.7){\line(1,0){0.5}}
\put(0.5,0.7){\line(1,1){0.5}}
\put(0.5,0.7){\line(1,-1){0.5}}
\put(1.5,1.2){\line(-1,-2){0.5}}
\put(1.0,1.2){\line(1,-2){0.18}}
\put(1.5,0.2){\line(-1,2){0.18}}
\put(1,1.2){\line(1,0){1}}
\put(1,0.2){\line(1,0){1}}
\put(0.25,0.9){\makebox(0,0)[b]{$#1$}}
\put(2.05,1.2){\makebox(0,0)[l]{$#2$}}
\put(2.05,0.2){\makebox(0,0)[l]{$#3$}}
\end{picture}
}}
\hfill}

\newcommand{\boxxbmcrossxNLO}[4]{
\mbox{\parbox{3.5cm}{\hspace{0.25cm}
\begin{picture}(3,1.4)
\thicklines
\put(0.7,0.2){\vector(-1,0){0.1}}
\put(2.3,0.2){\vector(1,0){0.1}}
\put(2.3,1.2){\vector(1,0){0.1}}
\put(0.8,1.2){\vector(1,0){0.1}}
\put(0.5,0.2){\line(1,0){2}}
\put(1.5,0.2){\line(1,2){0.2}}
\put(2,1.2){\line(-1,-2){0.2}}
\put(0.5,1.2){\line(1,0){2}}
\put(1,1.2){\line(1,-2){0.5}}
\put(1.5,1.2){\line(1,-2){0.5}}
\put(0.45,1.2){\makebox(0,0)[r]{$#1$}}
\put(0.45,0.2){\makebox(0,0)[r]{$#2$}}
\put(2.55,1.2){\makebox(0,0)[l]{$#3$}}
\put(2.55,0.2){\makebox(0,0)[l]{$#4$}}
\end{picture}
}} 
\hfill}

\newcommand{\boxxbmcrossxNLOtwo}[4]{
\mbox{\parbox{3.5cm}{\hspace{0.25cm}
\begin{picture}(3,1.4)
\thicklines
\put(0.7,0.2){\vector(-1,0){0.1}}
\put(2.3,0.2){\vector(1,0){0.1}}
\put(2.3,1.2){\vector(1,0){0.1}}
\put(0.8,1.2){\vector(1,0){0.1}}
\put(0.5,0.2){\line(1,0){2}}
\put(1.5,0.2){\line(1,2){0.2}}
\put(2,1.2){\line(-1,-2){0.2}}
\put(0.5,1.2){\line(1,0){2}}
\put(1,1.2){\line(1,-2){0.5}}
\put(1.5,1.2){\line(1,-2){0.5}}
\put(0.45,1.2){\makebox(0,0)[r]{$#1$}}
\put(0.45,0.2){\makebox(0,0)[r]{$#2$}}
\put(2.55,1.2){\makebox(0,0)[l]{$#3$}}
\put(2.55,0.2){\makebox(0,0)[l]{$#4$}}
\put(1.15,1.0){\makebox(0,0)[l]{$_{(2)}$}}
\end{picture}
}} 
\hfill}

\newcommand{\doublecrossNLO}[4]{
\mbox{\parbox{3.5cm}{\hspace{0.25cm}
\begin{picture}(3,1.4)
\thicklines
\put(0.7,0.2){\vector(-1,0){0.1}}
\put(2.3,0.2){\vector(1,0){0.1}}
\put(2.3,1.2){\vector(1,0){0.1}}
\put(0.8,1.2){\vector(1,0){0.1}}
\put(0.5,0.2){\line(1,0){2}}
\put(1.5,0.2){\line(1,2){0.2}}
\put(2,1.2){\line(-1,-2){0.2}}
\put(0.5,1.2){\line(1,0){2}}
\put(1,0.2){\line(0,1){1}}
\put(1.5,1.2){\line(1,-2){0.5}}
\put(0.45,1.2){\makebox(0,0)[r]{$#1$}}
\put(0.45,0.2){\makebox(0,0)[r]{$#2$}}
\put(2.55,1.2){\makebox(0,0)[l]{$#3$}}
\put(2.55,0.2){\makebox(0,0)[l]{$#4$}}
\end{picture}
}} 
\hfill}

\newcommand{\doublecrossNLOtwo}[4]{
\mbox{\parbox{3.5cm}{\hspace{0.25cm}
\begin{picture}(3,1.4)
\thicklines
\put(0.7,0.2){\vector(-1,0){0.1}}
\put(2.3,0.2){\vector(1,0){0.1}}
\put(2.3,1.2){\vector(1,0){0.1}}
\put(0.8,1.2){\vector(1,0){0.1}}
\put(0.5,0.2){\line(1,0){2}}
\put(1.5,0.2){\line(1,2){0.2}}
\put(2,1.2){\line(-1,-2){0.2}}
\put(0.5,1.2){\line(1,0){2}}
\put(1,0.2){\line(0,1){1}}
\put(1.5,1.2){\line(1,-2){0.5}}
\put(0.45,1.2){\makebox(0,0)[r]{$#1$}}
\put(0.45,0.2){\makebox(0,0)[r]{$#2$}}
\put(2.55,1.2){\makebox(0,0)[l]{$#3$}}
\put(2.55,0.2){\makebox(0,0)[l]{$#4$}}
\put(1.075,1.0){\makebox(0,0)[l]{$_{(2)}$}}
\end{picture}
}} 
\hfill}

\newcommand{\doublecrossxNLO}[4]{
\mbox{\parbox{3.5cm}{\hspace{0.25cm}
\begin{picture}(3,1.4)
\thicklines
\put(0.7,0.2){\vector(-1,0){0.1}}
\put(2.3,0.2){\vector(1,0){0.1}}
\put(2.3,1.2){\vector(1,0){0.1}}
\put(0.8,1.2){\vector(1,0){0.1}}
\put(0.5,0.2){\line(1,0){2}}
\put(1,0.2){\line(1,2){0.2}}
\put(1.5,1.2){\line(-1,-2){0.2}}
\put(0.5,1.2){\line(1,0){2}}
\put(2,0.2){\line(0,1){1}}
\put(1,1.2){\line(1,-2){0.5}}
\put(0.45,1.2){\makebox(0,0)[r]{$#1$}}
\put(0.45,0.2){\makebox(0,0)[r]{$#2$}}
\put(2.55,1.2){\makebox(0,0)[l]{$#3$}}
\put(2.55,0.2){\makebox(0,0)[l]{$#4$}}
\end{picture}
}} 
\hfill}

\newcommand{\doublecrossxNLOtwo}[4]{
\mbox{\parbox{3.5cm}{\hspace{0.25cm}
\begin{picture}(3,1.4)
\thicklines
\put(0.7,0.2){\vector(-1,0){0.1}}
\put(2.3,0.2){\vector(1,0){0.1}}
\put(2.3,1.2){\vector(1,0){0.1}}
\put(0.8,1.2){\vector(1,0){0.1}}
\put(0.5,0.2){\line(1,0){2}}
\put(1,0.2){\line(1,2){0.2}}
\put(1.5,1.2){\line(-1,-2){0.2}}
\put(0.5,1.2){\line(1,0){2}}
\put(2,0.2){\line(0,1){1}}
\put(1,1.2){\line(1,-2){0.5}}
\put(0.45,1.2){\makebox(0,0)[r]{$#1$}}
\put(0.45,0.2){\makebox(0,0)[r]{$#2$}}
\put(2.55,1.2){\makebox(0,0)[l]{$#3$}}
\put(2.55,0.2){\makebox(0,0)[l]{$#4$}}
\put(1.575,1.0){\makebox(0,0)[l]{$_{(3)}$}}
\end{picture}
}} 
\hfill}

\begin{document}
\unitlength1cm
\begin{titlepage}
\vspace*{-1cm}
\begin{flushright}
CERN-TH/2001-005\\
hep-ph/0101124\\
January 2001 
\end{flushright}                                
\vskip 3.5cm

\begin{center}
\boldmath
{\Large\bf Two-Loop Master Integrals for $\gamma^* \to 3$ Jets: \\[3mm]
the Non-Planar Topologies }\unboldmath
\vskip 1.cm
{\large  T.~Gehrmann}$^a$ and {\large E.~Remiddi}$^b$ 
\vskip .7cm
{\it $^a$ Theory Division, CERN, CH-1211 Geneva 23, Switzerland}
\vskip .4cm
{\it $^b$ Dipartimento di Fisica,
    Universit\`{a} di Bologna and INFN, Sezione di 
    Bologna,  I-40126 Bologna, Italy} 
\end{center}
\vskip 2.6cm

\begin{abstract}
The calculation of the two-loop corrections to the three-jet production 
rate and to event shapes in electron--positron annihilation requires 
the computation of a number of two-loop four-point 
master 
integrals with one off-shell and three on-shell legs. 
Up to now, only those master integrals corresponding to planar 
topologies were known. In this paper,
we compute the yet outstanding non-planar master integrals 
by solving differential equations in the external 
invariants which are fulfilled  by these master integrals. 
We obtain the master integrals 
 as expansions in $\e=(4-d)/2$, where $d$ is the
space-time dimension. The fully analytic results are expressed in terms of 
the two-dimensional harmonic polylogarithms already introduced in the 
evaluation of the planar topologies. 
\end{abstract}
\vfill
\end{titlepage}                                                                
\newpage

\renewcommand{\theequation}{\mbox{\arabic{section}.\arabic{equation}}}

\section{Introduction}
\setcounter{equation}{0}

Precision applications of particle physics phenomenology often demand 
theoretical predictions at the next-to-next-to-leading order 
in perturbation theory. Corrections to this order are known
for many inclusive observables, such as total cross sections or sum rules.
For  $2\to 2$ scattering and $1\to 3$ decay 
processes, the calculation of next-to-next-to-leading order corrections 
is a yet outstanding task. One of the major ingredients for these 
calculations consists the two-loop virtual corrections to the 
corresponding four-point Feynman amplitudes. 

Using dimensional regularization~\cite{dreg1,dreg2,hv} 
with $d=4-2\e$ dimensions as regulator for
ultraviolet and infrared divergences, 
the large number of different integrals appearing in the 
two-loop Feynman amplitudes for $2\to 2$ scattering or $1\to 3$ decay 
processes 
can be reduced to a small number of master integrals. 
The techniques used in these reductions are 
integration-by-parts identities~\cite{hv,chet1,chet2} 
and Lorentz invariance~\cite{gr}. A computer algorithm for the 
automatic reduction of all two-loop four-point integrals 
was described in~\cite{gr}. 

For two-loop four-point functions with massless internal propagators
and  all legs on-shell, which are 
relevant for example to the next-to-next-to-leading order
calculation of two-jet production at hadron colliders, all master
integrals have been calculated over the past
year~\cite{onshell1,onshell2,onshell3,onshell4,onshell5,onshell6}.
Very recently, these master integrals were already applied in the 
calculation of 
two-loop virtual corrections to Bhabha scattering~\cite{m1}
 in the limit of vanishing 
electron mass and to quark--quark scattering~\cite{m2}.

A different class of master integrals is required for the 
computation of next-to-next-to-leading order corrections to 
observables such as the three-jet production rate in electron--positron
annihilation, two-plus-one-jet production in deep inelastic
electron--proton scattering or vector-boson-plus-jet production at hadron 
colliders: two-loop
four-point functions with massless internal propagators and 
 one external leg off-shell. These functions involve one more scale than 
their on-shell counterparts, and are therefore more complicated and also
more numerous. Up to now, only the master integrals for planar 
topologies~\cite{smirnew,grplanar} as well as one of the 
non-planar master integrals were known~\cite{smirnp}. In~\cite{smirnew,smirnp},
Smirnov used a Mellin--Barnes countour integral technique to derive 
one planar and one non-planar master integral, obtaining analytic expressions 
for the divergent parts and an integral representation (one-dimensional 
in the planar case and three-dimensional in the non-planar case) 
for the finite part. Using the differential equation technique 
derived in~\cite{gr}, we computed the full set of planar master integrals 
in~\cite{grplanar}. Complementary work on 
the purely numerical evaluation of this 
type of master integrals has been presented recently by Binoth and 
Heinrich~\cite{num}.

Our results
for the planar topologies~\cite{grplanar} 
were given in fully analytic form, in terms of a 
new class of functions, two-dimensional harmonic polylogarithms
(2dHPL). 
The 2dHPL are an extension of the harmonic 
polylogarithms (HPL) of~\cite{hpl}. All HPL and 2dHPL that appear in the 
divergent parts of the planar master integrals have weight 
$\leq 3$ and can be 
related to the more commonly known Nielsen's generalized 
polylogarithms~\cite{nielsen,bit} of suitable arguments. 
The functions of weight 4 appearing in the finite 
parts of the master integrals can all be represented, by the very 
definition, as one-dimensional 
integrals over Nielsen's polylogarithms of weight 3, hence of Nielsen's 
generalized polylogarithms of suitable arguments according to the above 
remark. A table with all relations 
is included in the appendix of~\cite{grplanar}. 

It is the purpose of  the  present paper to extend our work of~\cite{grplanar}
towards the computation of  
non-planar four-point master integrals with one off-shell leg from their 
differential equations. In Section~\ref{sec:de}, we discuss the use of 
the differential equation method in the computation of master integrals.
We point out in particular the differences to the 
calculation of the planar master integrals. 
We then list our 
analytical results for the complete set of non-planar master integrals 
in 
Section~\ref{sec:mi}. Our results are, like in the planar case, expressed in 
terms of 2dHPL.
In Section~\ref{sec:mi}  we also compare our results
with the existing partial ones
available in the literature.
Finally, Section~\ref{sec:conc} contains conclusions and an outlook on 
future applications. 

\section{Solving Differential Equations for Master Integrals}
\label{sec:de}
\setcounter{equation}{0}

The use of differential equations for the computation of master integrals 
was first suggested in~\cite{kotikov} as a means of relating integrals 
with massive internal propagators to their massless counterparts. The
method was developed in detail in~\cite{remiddi}, where it was also 
extended towards differential equations in the external invariants. 
As a first application of this method, the two-loop sunrise diagram 
with arbitrary internal masses was studied in~\cite{appl}. We 
have developed the differential equation formalism for
two-loop four-point functions with massless internal propagators,
three external legs on-shell and
one external leg off-shell in~\cite{gr}.
We derived an algorithm for the automatic reduction, 
by means of computer algebra (using FORM~\cite{form} and Maple~\cite{maple}),
of any
two-loop four-point integral to a small set of master integrals, for which we 
derived differential equations in the external invariants.

Four-point functions depend on three linearly independent momenta:
$p_1$, $p_2$ and $p_3$. In our calculation, we take all these momenta 
on-shell ($p_i^2=0$), 
while the fourth momentum $p_{123}=(p_1+p_2+p_3)$ is taken
off-shell. The kinematics of the four-point function is then 
fully described by specifying the values of the three Lorentz 
invariants $\sab = (p_1+p_2)^2$, $\sac = (p_1+p_3)^2$ and $\sbc =
(p_2+p_3)^2$. 

Expressing the system of differential equations obtained in~\cite{gr} for 
any master integral in the variables 
$\sabc = \sab + \sac + \sbc$, $y=\sac/\sabc$ and $z=\sbc/\sabc$, we
obtain a homogeneous equation in $\sabc$, and inhomogeneous equations 
in $y$ and $z$. Since $\sabc$ is the only quantity carrying a mass 
dimension, 
the corresponding differential equation is nothing but the rescaling 
relation obtained by investigating the behaviour of the master integral 
under a rescaling of all external momenta by a constant factor. 

In the reduction of non-planar two-loop four-point functions 
with one off-shell leg, one obtains (in addition to master integrals 
corresponding to planar subtopologies) several genuinely 
non-planar master integrals, corresponding to three- and 
four-point topologies. The two master integrals corresponding to 
three-point topologies are the crossed vertex integral with one leg off-shell, 
two legs on-shell and the crossed vertex integral with two legs off-shell, 
one leg on-shell. The former obeys only
a homogeneous differential equation, 
which is of no use for its complete computation, but it is otherwise 
relatively simple and already known in the literature~\cite{kl1,kl2}. We
list its value in Section~\ref{sec:mi} for reference. 
For the latter vertex 
integral with two off-shell legs, one obtains one homogeneous and one 
inhomogeneous differential equation. The computation of this integral 
from its inhomogeneous differential equation is straightforward, following 
the procedure used for the planar integrals~\cite{grplanar}.

Of the four non-planar four-point topologies with one off-shell leg, 
only one is reducible to planar subtopologies; the remaining three 
contain two master integrals each~\cite{gr}. 
For the six-propagator
topology, we use the following basis of master integrals:
\begin{eqnarray}
\boxxbmcrossxNLO{p_{123}}{p_3}{p_2}{p_1} &=& \int \frac{\d^d
k}{(2\pi)^d}\frac{\d^d l}{(2\pi)^d}  \frac{1}{k^2 (k-p_{123})^2 
(k-l)^2 (k-l-p_1)^2 l^2 (l-p_2)^2 },
\label{eq:sixa}
\\ 
\boxxbmcrossxNLOtwo{p_{123}}{p_3}{p_2}{p_1} &=& \int \frac{\d^d
k}{(2\pi)^d}\frac{\d^d l}{(2\pi)^d}  \frac{2 k\cdot p_2}{k^2 (k-p_{123})^2 
(k-l)^2 (k-l-p_1)^2 l^2 (l-p_2)^2 }.
\label{eq:sixb}
\end{eqnarray}
For the two seven-propagator  topologies, we choose: 
\begin{eqnarray}
\doublecrossNLO{p_{123}}{p_3}{p_2}{p_1}\hspace{-4mm}
 &=& \int \frac{\d^d
k}{(2\pi)^d}\frac{\d^d l}{(2\pi)^d}  \frac{1}{k^2 (k-p_{12})^2 (k-p_{123})^2 
(k-l)^2 (k-l-p_1)^2 l^2 (l-p_2)^2 },
\label{eq:sevena}
\\ 
\doublecrossNLOtwo{p_{123}}{p_3}{p_2}{p_1}\hspace{-4mm} &=& \int \frac{\d^d
k}{(2\pi)^d}\frac{\d^d l}{(2\pi)^d}  \frac{2 k\cdot p_2}{k^2 (k-p_{12})^2
(k-p_{123})^2 
(k-l)^2 (k-l-p_1)^2 l^2 (l-p_2)^2 },
\label{eq:sevenb}
\end{eqnarray}
and,
\begin{eqnarray}
\doublecrossxNLO{p_{123}}{p_3}{p_2}{p_1}\hspace{-4mm} &=& \int \frac{\d^d
k}{(2\pi)^d}\frac{\d^d l}{(2\pi)^d}  \frac{1}{k^2  (k-p_{123})^2 
(k-l)^2 (k-l-p_3)^2 l^2 (l-p_2)^2 (l-p_{12})^2},
\label{eq:sevenc}
\\ 
\doublecrossxNLOtwo{p_{123}}{p_3}{p_2}{p_1}\hspace{-4mm} &=& \int \frac{\d^d
k}{(2\pi)^d}\frac{\d^d l}{(2\pi)^d}  \frac{2 l\cdot p_3}{k^2 (k-p_{123})^2 
(k-l)^2 (k-l-p_3)^2 l^2 (l-p_2)^2 (l-p_{12})^2}.
\label{eq:sevend}
\end{eqnarray}
This choice of basis is not unambiguous. Our main motivation for this 
particular basis is the decoupling, in the $ d \to 4 $ limit, of the 
homogeneous part of the differential equations 
for each topology, which is crucial for the applicability of our 
algorithm to the solution of the differential equations outlined below. 

Quite in general, the analytic expression of all planar and non-planar 
amplitudes is found to consist of several terms, each equal to a 
prefactor (a simple rational polynomial in the external invariants) 
times a combination of 2dHPL multiplied by simple coefficients 
(integer numbers or rational coefficients). 
The major difference between the planar and non-planar master integrals 
is in the number of such terms. Indeed, all planar master 
integrals can be written as the product of a single prefactor 
times a Laurent series in $\e = (4-d)/2 $ containing 
as coefficients only 
logarithms and polylogarithms of ratios of the invariants. It has become 
clear already in the on-shell case~\cite{onshell2} that the non-planar
master integrals do not take such a simple form. This expectation was 
confirmed for the off-shell non-planar master integrals in~\cite{num},
where the $1/\e^4$ and $1/\e^3$ terms of (\ref{eq:sevena}) and 
(\ref{eq:sevenc}) were calculated analytically. All these non-planar integrals 
contain the sum of several different prefactors, each 
multiplied by a different Laurent series in $\e$. 

The algorithm employed in our calculation of the planar master integrals 
from their differential equations~\cite{grplanar} relied on the 
possibility of determining a unique rational prefactor for each master 
integral. Once this prefactor was found, the coefficients in the 
Laurent series were determined by inserting a general ansatz containing 
all possible harmonic polylogarithms, up to a given weight, and a subsequent 
matching of the individual terms. This procedure has to be modified to 
account for the more involved structure of the non-planar master 
integrals.

With two master integrals $T_1(y,z,\e)$ and $T_2(y,z,\e)$ 
for a given topology, the 
differential equations in $y$ take the following form, exact in $\e$: 
\begin{eqnarray}
\frac{\partial}{\partial y} T_1(y,z,\e) 
&=& R_{11}(y,z,\e) T_1(y,z,\e) + R_{12}(y,z,\e) T_2(y,z,\e) + I_1(y,z,\e) 
\;,\nonumber \\
\frac{\partial}{\partial y} T_2(y,z,\e) &=& R_{21}(y,z,\e) T_1(y,z,\e)
 + R_{22}(y,z,\e) T_2(y,z,\e) + I_2 (y,z,\e)\;,
\label{eq:difsyst}
\end{eqnarray}
where the $R_{ij}(y,z,\e)$ are rational functions of $y,z$ and $\e$, 
of order 1 in $\e$;  $I_1(y,z,\e)$ and 
$I_2(y,z,\e)$ are the inhomogeneous terms, containing the simpler and known 
master integrals of subtopologies 
of the topology under consideration.
The first master integral $T_1(y,z,\e)$ is always chosen to contain no 
scalar products in the numerator and all propagators only to unit power 
in the denominator, see 
(\ref{eq:sixa}),(\ref{eq:sevena}),(\ref{eq:sevenc}). 
The second master integral $T_2(y,z,\e)$ can be chosen  
to contain one or more irreducible scalar products in the numerator, or 
one or more squared propagators in the denominator, or even a combination 
of both. The only requirement on $T_2(y,z,\e)$ is its independence 
of $T_1(y,z,\e)$ under 
integration-by-parts and Lorentz invariance identities. Any integral of 
the topology under consideration can then be expressed as a linear combination
of $T_1$ and $T_2$ plus simpler subtopologies. 
By exploiting the freedom to choose $T_2(y,z,\e)$, we always succeeded 
in finding a $T_2(y,z,\e)$ for which $R_{21}(y,z,\e)$ in (\ref{eq:difsyst}) 
is of order $\e$ (this condition is fulfilled for the basis of master 
integrals (\ref{eq:sixa})--(\ref{eq:sevend}) introduced above). 
As a consequence, the $y$-differential equation for $T_2(y,z,\e)$, 
when Laurent-expanded in $\e$, decouples from $T_1(y,z,\e)$ and the 
coefficients of the expansions can be obtained from the differential 
equations by recursive quadrature. 

Let us illustrate the algorithm when $T_1(y,z,\e)$ and $T_2(y,z,\e)$ are 
the amplitudes corresponding to (\ref{eq:sevenc}) and (\ref{eq:sevend}), 
respectively. The explicit form of the equations, with $d=4-2\e$,  
in that case is 
\begin{eqnarray} 
\frac{\partial}{\partial y} T_1(y,z,\e) 
&=& \phantom{+} 
    \left( - \frac{d-4}{y} + \frac{d-5}{y+z} + \frac{1}{1-y-z} 
    \right) T_1(y,z,\e) \nonumber\\ 
&& + \frac{2d-9}{z} \left( \frac{1}{y} - \frac{1}{y+z} \right) T_2(y,z,\e) 
                 + I_1(y,z,\e) \ ; \nonumber\\ 
\frac{\partial}{\partial y} T_2(y,z,\e) 
&=& -\; (d-4)\; \frac{z}{2y} T_1(y,z,\e) \nonumber\\ 
&&  + \left( \frac{3d-14}{2y} + \frac{1}{1-y-z} \right) T_2(y,z,\e) 
                 + I_2(y,z,\e) \ . 
\label{eq:exex} 
\end{eqnarray} 
We now expand everything in Laurent series in $\e$, up to terms of 
order 0, as 
\begin{equation}
 T_k(y,z,\e) = \sum_{i=0}^4 \frac{1}{\e^i} T_{k,i}(y,z)\ , 
   \hspace{1.8cm} 
   I_k(y,z,\e) = \sum_{i=0}^4 \frac{1}{\e^i} I_{k,i}(y,z)\ ; 
\end{equation}
the equations then read 
\begin{eqnarray} 
\frac{\partial}{\partial y} T_{1,i}(y,z) 
&=& \phantom{+} 
    \left( \frac{1}{1-y-z} - \frac{1}{y+z} \right) T_{1,i}(y,z) \nonumber\\ 
&& + \frac{1}{z}\left( - \frac{1}{y} + \frac{1}{y+z} \right) T_{2,i}(y,z) 
   + J_{1,i}(y,z)\ , \nonumber\\ 
\frac{\partial}{\partial y} T_{2,i}(y,z) 
&=& \left( - \frac{1}{y} + \frac{1}{1-y-z} \right) T_{2,i}(y,z) 
   + J_{2,i}(y,z)\ , 
\label{eq:exex1} 
\end{eqnarray} 
where the new inhomogeneous terms are 
\begin{eqnarray} 
J_{1,i}(y,z) &=& \left( - \frac{1}{y} + \frac{1}{y+z} 
                 \right) T_{1,i+1}(y,z) 
                 + \frac{2}{z}\; \left( \frac{1}{y} - \frac{1}{y+z} 
                 \right) T_{2,i+1}(y,z) + I_{1,i}(y,z) \ , \nonumber\\ 
J_{2,i}(y,z) &=& - \frac{z}{2y} T_{1,i+1}(y,z) 
                 + \frac{3}{2y} T_{2,i+1}(y,z) + I_{2,i}(y,z) \ ; 
\end{eqnarray} 
and it is understood that quantities whose second index is greater than 4 
vanish, as the greatest singularity of the Laurent expansion in $\e$ is 
4, according to the analysis of the 
soft and collinear structure~\cite{catani}. 
The system is solved bottom up by starting from the coefficients 
of greatest singularity corresponding to $i=4$, $T_{k,4}(y,z)$, 
whose equations involve as inhomogeneous terms only the 
$I_{k,4}(y,z)$, which are known. One then proceeds to the equations 
for $T_{k,3}(y,z)$, which involve as inhomogeneous terms  
$T_{k,4}(y,z)$ and $I_{k,3}(y,z)$, which are known by now
and so on until 
the $T_{k,0}(y,z)$ are obtained. 

According to the preceding general remarks, the equation for 
$ T_{2,i}(y,z) $ does not involve $ T_{1,i}(y,z) $ (it involves 
$ T_{1,i+1}(y,z) $, which is known when considering the $i$-th term 
of the Laurent expansion); the homogeneous part of the equation is 
almost trivial, having as a solution $ 1/[y(1-y-z)] $; 
by following the standard method of the variation of the constants, 
the solution of the inhomogeneous equation is immediately given by 
the quadrature formula 
\begin{equation} 
 T_{2,i}(y,z) = \frac{1}{y(1-y-z)} 
              \int^y \d y'\; y'\; (1-y'-z) J_{2,i}(y',z) \ . 
\label{eq:exex2} 
\end{equation} 
The above result can be inserted into the equation for $ T_{1,i}(y,z) $, 
which is then solved in the same manner; starting from $i=4$, the whole 
procedure can be iterated for a smaller value of $i$,  till $i=0$ is 
reached (one could in principle arrive at any other desired value of $i$). 
Note that the whole process involves only the $y$-dependent denominators 
appearing in the differential equations in the variable $y$, 
namely $1/y$, $1/(1-y)$, 
$1/(y+z)$ and $1/(1-y-z)$. As the homogeneous terms are also equal to 
products of those denominators and 2dHPL functions, 
after careful partial fractioning of the $y$-dependent factors 
the integrations in (\ref{eq:exex2}) can always be evaluated 
analytically: when a square denominator, say $1/y^2$, occurs, they 
can be carried out almost trivially by integrating by parts, 
while when the integrand is a single denominator times any 2dHPL function 
of some weight $w$, the primitive is nothing but another 2dHPL function 
of weight $w+1$ (see the Appendix of~\cite{grplanar} for more details 
on 2dHPL). 

The quadrature formula (\ref{eq:exex2}) is definite up to a constant of 
integration in $y$, so one still needs 
one constant per master integral per order in $\e$. 
 These constants of integration (which are functions of $z$)
can be determined only by evaluating the master integrals for some boundary 
condition. In the planar case~\cite{grplanar}, we have used the fact that 
all planar master integrals, as well as their derivatives, 
are regular in the whole kinematical plane, with the exception of two branch 
points at $y=0$ and $z=0$. As a consequence, any factor $(1-y)$, $(1-y-z)$, or
$(y+z)$ appearing in the denominator of the
homogeneous term of the differential equations 
for a master integral could be used to determine the boundary condition in 
$y=1$, $y=(1-z)$, or $y=-z$, respectively: multiplying the 
differential equation with one of these factors and taking the limit where the 
factor vanishes, the derivative and all the other terms not having that 
factor as a denominator drop out and 
one obtains a linear relation expressing the master integral 
in this special kinematical point in terms of the known 
master integrals of 
its subtopologies. 
The non-planar master integrals possess a different analytic structure:
they have three branch points: in $y=0$, $z=0$ and $y=1-z$. Consequently,
only factors $(1-y)$ or $(y+z)$ in the denominator of the homogeneous term 
can be used for the determination of the boundary conditions. It turns out 
that these boundary conditions alone are not sufficient for a complete 
determination of the constants of integration in all the master integrals: 
for each topology,
one finds that one constant of integration per order in $\e$ remains 
unconstrained. We determine this 
($z$-dependent) constant by solving the system of 
$z$-differential equations for each topology
at the point $y=1$ and requiring this solution to be 
regular at the point $z=1$. Some care has to be taken in 
the analytic continuation of the master integrals across the 
cut in $y=1-z$ 
to the point $y=1$; details on the analytic continuation of the 2dHPL can be
found in the appendix of~\cite{grplanar}.

Using the integration procedure described here, we have determined
 all non-planar two-loop four-point master integrals with
one off-shell leg. The results are summarized in the following section.

\section{Master Integrals}
\label{sec:mi}
\setcounter{equation}{0}

In the reduction of a non-planar two-loop four-point function
 with one off-shell leg, one will in general encounter master integrals 
with non-planar topology in combination with master integrals 
corresponding to planar subtopologies. These planar master integrals were
all computed in~\cite{grplanar}. 
In this section, we tabulate all genuinely non-planar 
master integrals relevant to the
computation of 
two-loop four-point functions with one off-shell leg. 
We classify the integrals, according to the number of different 
kinematical scales on 
which they depend, into one-scale integrals, two-scale integrals and 
three-scale integrals. 

The common normalization factor of all master integrals is 
\begin{equation}
S_\e = \left[(4\pi)^\e \frac{ \Gamma (1+\e)
    \Gamma^2 (1-\e)}{ \Gamma (1-2\e)} \right] \; .
\end{equation}

\subsection{One-Scale Integrals}
\label{sec:mi1}

The only non-planar one-scale integral is the 
crossed on-shell vertex integral, which 
 was first  computed in~\cite{kl1,kl2}. We list the result only for 
completeness:
\begin{eqnarray}
\trianglecrossNLO{p_{12}}{p_1}{p_2}  &=&
 \left(\frac{S_\e}{16 \pi^2}\right)^2\,
 \left( -s_{12} \right)^{-2-2\e}\,
\left[ -\frac{1}{\e^4} + 
\frac{5\pi^2}{6\e^2} + \frac{23}{\e}\zeta_3 + \frac{103\pi^4}{180} 
+ {\cal O}(\e) \right]\;.
\end{eqnarray}

\subsection{Two-Scale Integrals}
\label{sec:mi2}
The crossed vertex integral with two off-shell legs is the only 
non-planar two-scale master integral. 
This integral was, to the best of our 
knowledge, not known up to now. 
It fulfils one inhomogeneous 
differential equation in the ratio of the two external invariants, which can 
be employed for its computation. We obtain:
\begin{equation}
\trianglecrossNLO{p_{123}}{p_{12}}{p_{3}} =  \left(\frac{S_\e}{16
\pi^2}\right)^2\, \frac{\left( -\sabc \right)^{-2\e}}{(\sabc-\sab)^2} \,
\sum_{i=0}^4 \frac{g_{6.1,i}
\left(\frac{\sab}{\sabc}\right)}{\e^i}\; + 
{\cal O}(\e) ,
\end{equation}
with:
\begin{eqnarray}
g_{6.1,4}(x) &=& 0 \ ,\\
g_{6.1,3}(x) &=& 0 \ , \\
g_{6.1,2}(x) &=& - 2 H(0,0;x)\ , \\
g_{6.1,1}(x) &=&           
          + 6 H(0,0,0;x)
          + 4 H(0,1,0;x)
          - 4 H(1,0,0;x)
          + \frac{2}{3} \pi^2 H(0;x)
          + 12 \zeta_3 \ , \\
g_{6.1,0}(x) &=&            
          - 14 H(0,0,0,0;x)
          - 2 H(0,0,1,0;x)
          - 6 H(0,1,0,0;x)
          + 4 H(0,1,1,0;x) \nonumber \\
&&
          + 12 H(1,0,0,0;x)
          + 8 H(1,0,1,0;x)
          - 8 H(1,1,0,0;x)
          + \frac{23\pi^4}{90}\nonumber \\
&& 
          + \zeta_3 \left[ -2 H(0;x) + 24 H(1;x) \right]
          + \frac{\pi^2}{6} \left[
          - 2 H(0,0;x)
          + 4 H(0,1;x)
          + 8 H(1,0;x)
                            \right] \ .
\end{eqnarray}

\subsection{Three-Scale Integrals}
\label{sec:mi3}

There are four different topologies of 
non-planar two-loop four-point functions with one off-shell leg, two 
topologies with six different  propagators, and two others with
seven different propagators. Only one of the six-propagator topologies
is fully reducible to simpler (planar) subtopologies~\cite{gr}. 
All other topologies contain two master integrals each. The two 
master integrals for each topology fulfil a coupled system of inhomogeneous 
differential equations in $y=\sac/\sabc$ and another coupled system in 
$z=\sbc/\sabc$. 
As explained in Section~\ref{sec:de}, solving 
one of these systems is already sufficient to 
compute the master integrals up to a constant of integration. The 
second system is used to constrain the boundary conditions and 
hence the constants of integration, as well as a check on the result.  

The choice of the basis of master integrals for each topology is described in
Section~\ref{sec:de}. In particular, we always choose the integral without 
scalar products in the numerator and no squared propagator as one 
of the master integrals. This integral is symmetric under the interchange of 
two external momenta, which we take to be $p_1$ and $p_2$.

The two master integrals for the six-propagator topology are:
\begin{equation}
\boxxbmcrossxNLO{p_{123}}{p_3}{p_2}{p_1} =\left(\frac{S_\e}{16
\pi^2}\right)^2\, \frac{\left( -\sabc \right)^{-2\e}}
{\sab\sabc} \sum_{i=0}^4 \frac{f_{6.2,i}
\left(\frac{\sac}{\sabc},\frac{\sbc}{\sabc}\right)}{\e^i}\; + 
{\cal O}(\e) ,
\end{equation}
with:
\begin{eqnarray}
f_{6.2,4}(y,z) &=& -\frac{1}{4}\ , \\
f_{6.2,3}(y,z) &=&           
          - \frac{1}{2} H(0;y)
          - \frac{1}{2} H(0;z)
          - \frac{1}{2} H(1;z)
          - \frac{1}{2} H(1-z;y)\ , \\
f_{6.2,2}(y,z) &=&  
          - H(0;y) H(0;z)
          - H(0;y) H(1;z)
          - H(0;z) H(1-z;y)
          + H(0,0;y)
          + H(0,0;z) \nonumber \\
&&
          - H(0,1;z)
          - H(0,1-z;y)
          - H(1;z) H(1-z;y)
          + 2 H(1,0;y)
          + H(1,0;z)\nonumber \\
&&
          - H(1,1;z)
          - H(1-z,0;y)
          - H(1-z,1-z;y)
          +\frac{\pi^2}{2}\ ,\\
f_{6.2,1}(y,z) &=&           + 2 H(0;y) H(1,0;z)
          - 2 H(0;y) H(1,1;z)
          + 4 H(0;z) H(1,0;y)
          - 2 H(0;z) H(1-z,0;y)\nonumber \\
&&
          - 2 H(0;z) H(1-z,1-z;y)
          + 2 H(0,0;y) H(0;z)
          + 2 H(0,0;y) H(1;z)
          + 2 H(0,0;z) H(0;y)\nonumber \\
&&
          + 2 H(0,0;z) H(1-z;y)
          - 2 H(0,0,0;y)
          - 2 H(0,0,0;z)
          + 2 H(0,0,1;z)
          + 2 H(0,0,1-z;y)\nonumber \\
&&
          - 2 H(0,1;z) H(0;y)
          - 2 H(0,1;z) H(1-z;y)
          + H(0,1,0;y)
          + 3 H(0,1,0;z)
          - 2 H(0,1,1;z)\nonumber \\
&&
          - 2 H(0,1-z;y) H(0;z)
          - 2 H(0,1-z;y) H(1;z)
          + 2 H(0,1-z,0;y)\nonumber \\
&&
          - 2 H(0,1-z,1-z;y)
          - 2 H(1;z) H(1-z,0;y)
          - 2 H(1;z) H(1-z,1-z;y)\nonumber \\
&&
          + 4 H(1,0;y) H(1;z)
          + 2 H(1,0;z) H(1-z;y)
          - 4 H(1,0,0;y)
          - 2 H(1,0,0;z)\nonumber \\
&&
          + 2 H(1,0,1;z)
          + 4 H(1,0,1-z;y)
          - 2 H(1,1;z) H(1-z;y)
          - 3 H(1,1,0;y)
          + 3 H(1,1,0;z)\nonumber \\
&&
          - 2 H(1,1,1;z)
          + 4 H(1,1-z,0;y)
          + 2 H(1-z,0,0;y)
          - 2 H(1-z,0,1-z;y)\nonumber \\
&&
          + 4 H(1-z,1,0;y)
          - 2 H(1-z,1-z,0;y)
          - 2 H(1-z,1-z,1-z;y)
          \nonumber \\
&&
          +\frac{23}{2}\zeta_3 
          + \frac{\pi^2}{6} \Big[
              + 3 H(0;y)
              + 3 H(0;z)
              - 3 H(1;y)
              + 3 H(1;z)
              + 6 H(1-z;y)    \Big] \ , \\
f_{6.2,0}(y,z) &=& 
          - 4 H(0;y) H(1,0,0;z)
          - 2 H(0;y) H(1,0,1;z)
          - 4 H(0;y) H(1,1,0;z)\nonumber \\
&&
          - 4 H(0;y) H(1,1,1;z)
          - 8 H(0;z) H(1,0,0;y)
          + 2 H(0;z) H(1,0,1-z;y)\nonumber \\
&&
          - 8 H(0;z) H(1,1-z,0;y)
          + 4 H(0;z) H(1-z,0,0;y)
          - 4 H(0;z) H(1-z,0,1-z;y)\nonumber \\
&&
          + 8 H(0;z) H(1-z,1,0;y)
          - 4 H(0;z) H(1-z,1-z,0;y)\nonumber \\
&&
          - 4 H(0;z) H(1-z,1-z,1-z;y)
          - 4 H(0,0;y) H(0,0;z)
          + 4 H(0,0;y) H(0,1;z)\nonumber \\
&&
          - 4 H(0,0;y) H(1,0;z)
          + 4 H(0,0;y) H(1,1;z)
          + 4 H(0,0;z) H(0,1-z;y)\nonumber \\
&&
          - 8 H(0,0;z) H(1,0;y)
          + 4 H(0,0;z) H(1-z,0;y)
          + 4 H(0,0;z) H(1-z,1-z;y)\nonumber \\
&&
          - 4 H(0,0,0;y) H(0;z)
          - 4 H(0,0,0;y) H(1;z)
          - 4 H(0,0,0;z) H(0;y)\nonumber \\
&&
          - 4 H(0,0,0;z) H(1-z;y)
          + 4 H(0,0,0,0;y)
          + 4 H(0,0,0,0;z)
          + 2 H(0,0,0,1;z)\nonumber \\
&&
          - 4 H(0,0,0,1-z;y)
          - 2 H(0,0,1;z) H(0;y)
          - 6 H(0,0,1;z) H(1;y)\nonumber \\
&&
          + 4 H(0,0,1;z) H(1-z;y)
          - 2 H(0,0,1,0;y)
          - 6 H(0,0,1,0;z)
          + 4 H(0,0,1,1;z)\nonumber \\
&&
          - 2 H(0,0,1-z;y) H(0;z)
          + 4 H(0,0,1-z;y) H(1;z)
          - 4 H(0,0,1-z,0;y)\nonumber \\
&&
          + 4 H(0,0,1-z,1-z;y)
          + 6 H(0,0,z;y) H(1;z)
          + 6 H(0,0,z,1-z;y)\nonumber \\
&&
          - 10 H(0,1;z) H(0,1-z;y)
          + 8 H(0,1;z) H(1,0;y)
          - 6 H(0,1;z) H(1,1-z;y)\nonumber \\
&&
          - 4 H(0,1;z) H(1-z,0;y)
          - 4 H(0,1;z) H(1-z,1-z;y)
          + 8 H(0,1,0;y) H(0;z)\nonumber \\
&&
          + 8 H(0,1,0;y) H(1;z)
          - 4 H(0,1,0;z) H(0;y)
          - 10 H(0,1,0;z) H(1;y)\nonumber \\
&&
          + 6 H(0,1,0;z) H(1-z;y)
          - 2 H(0,1,0,0;y)
          - 6 H(0,1,0,0;z)
          + 6 H(0,1,0,1;z)\nonumber \\
&&
          + 8 H(0,1,0,1-z;y)
          - 4 H(0,1,1;z) H(0;y)
          - 4 H(0,1,1;z) H(1-z;y)
          - 9 H(0,1,1,0;y)\nonumber \\
&&
          - 5 H(0,1,1,0;z)
          - 4 H(0,1,1,1;z)
          + 8 H(0,1,1-z,0;y)
          - 6 H(0,1-z;y) H(1,0;z)\nonumber \\
&&
          - 4 H(0,1-z;y) H(1,1;z)
          - 12 H(0,1-z,0;y) H(0;z)
          + 4 H(0,1-z,0;y) H(1;z)\nonumber \\
&&
          - 4 H(0,1-z,0,0;y)
          + 4 H(0,1-z,0,1-z;y)
          - 8 H(0,1-z,1,0;y)\nonumber \\
&&
          - 4 H(0,1-z,1-z;y) H(0;z)
          - 4 H(0,1-z,1-z;y) H(1;z)
          + 4 H(0,1-z,1-z,0;y)\nonumber \\
&&
          - 4 H(0,1-z,1-z,1-z;y)
          - 6 H(0,1-z,z;y) H(1;z)
          - 6 H(0,1-z,z,1-z;y)\nonumber \\
&&
          + 4 H(1;z) H(1-z,0,0;y)
          - 4 H(1;z) H(1-z,0,1-z;y)
          + 8 H(1;z) H(1-z,1,0;y)\nonumber \\
&&
          - 4 H(1;z) H(1-z,1-z,0;y)
          - 4 H(1;z) H(1-z,1-z,1-z;y)
          - 8 H(1,0;y) H(1,0;z)\nonumber \\
&&
          + 8 H(1,0;y) H(1,1;z)
          - 10 H(1,0;z) H(1,1-z;y)
          + 4 H(1,0;z) H(1-z,0;y)\nonumber \\
&&
          + 4 H(1,0;z) H(1-z,1-z;y)
          - 8 H(1,0,0;y) H(1;z)
          - 4 H(1,0,0;z) H(1-z;y)\nonumber \\
&&
          + 8 H(1,0,0,0;y)
          + 4 H(1,0,0,0;z)
          + 2 H(1,0,0,1;z)
          - 8 H(1,0,0,1-z;y)\nonumber \\
&&
          - 6 H(1,0,1;z) H(1;y)
          + 4 H(1,0,1;z) H(1-z;y)
          - 4 H(1,0,1,0;y)
          - 6 H(1,0,1,0;z)\nonumber \\
&&
          + 4 H(1,0,1,1;z)
          + 8 H(1,0,1-z;y) H(1;z)
          - 8 H(1,0,1-z,0;y)\nonumber \\
&&
          + 8 H(1,0,1-z,1-z;y)
          + 6 H(1,0,z;y) H(1;z)
          + 6 H(1,0,z,1-z;y)\nonumber \\
&&
          - 4 H(1,1;z) H(1-z,0;y)
          - 4 H(1,1;z) H(1-z,1-z;y)
          - 10 H(1,1,0;z) H(1;y)\nonumber \\
&&
          + 6 H(1,1,0;z) H(1-z;y)
          + 6 H(1,1,0,0;y)
          - 6 H(1,1,0,0;z)
          + 6 H(1,1,0,1;z)\nonumber \\
&&
          - 4 H(1,1,1;z) H(1-z;y)
          - 3 H(1,1,1,0;y)
          - 5 H(1,1,1,0;z)
          - 4 H(1,1,1,1;z)\nonumber \\
&&
          + 8 H(1,1-z,0;y) H(1;z)
          - 8 H(1,1-z,0,0;y)
          + 8 H(1,1-z,0,1-z;y)\nonumber \\
&&
          - 16 H(1,1-z,1,0;y)
          + 8 H(1,1-z,1-z,0;y)
          - 6 H(1,1-z,z;y) H(1;z)\nonumber \\
&&
          - 6 H(1,1-z,z,1-z;y)
          - 4 H(1-z,0,0,0;y)
          + 4 H(1-z,0,0,1-z;y)\nonumber \\
&&
          + 2 H(1-z,0,1,0;y)
          + 4 H(1-z,0,1-z,0;y)
          - 4 H(1-z,0,1-z,1-z;y)\nonumber \\
&&
          - 8 H(1-z,1,0,0;y)
          + 8 H(1-z,1,0,1-z;y)
          - 6 H(1-z,1,1,0;y)\nonumber \\
&&
          + 8 H(1-z,1,1-z,0;y)
          + 4 H(1-z,1-z,0,0;y)
          - 4 H(1-z,1-z,0,1-z;y)\nonumber \\
&&
          + 8 H(1-z,1-z,1,0;y)
          - 4 H(1-z,1-z,1-z,0;y)
          - 4 H(1-z,1-z,1-z,1-z;y)\nonumber \\
&&
          +\frac{31\pi^4}{360}
          + \zeta_3   \Big[
           + 10 H(0;y)
           + 10 H(0;z)
           - 13 H(1;y)
           + 10 H(1;z)
           + 23 H(1-z;y)\Big]\nonumber \\
&&
          + \frac{\pi^2}{6} \Big[
           - 4 H(0;y) H(0;z)
           - 4 H(0;y) H(1;z)
           - 10 H(0;z) H(1;y)
           + 6 H(0;z) H(1-z;y)\nonumber \\
&&
           - 6 H(0,0;y)
           - 6 H(0,0;z)
           - 9 H(0,1;y)
           - 7 H(0,1;z)
           + 2 H(0,1-z;y)
           - 10 H(1;y) H(1;z)\nonumber \\
&&
           + 6 H(1;z) H(1-z;y)
           - 12 H(1,0;y)
           - 6 H(1,0;z)
           - 3 H(1,1;y)
           - 7 H(1,1;z)\nonumber \\
&&
           - 10 H(1,1-z;y)
           + 6 H(1-z,0;y)
           - 6 H(1-z,1;y)
           + 12 H(1-z,1-z;y) \Big] \ .
\end{eqnarray}

\begin{equation}
\boxxbmcrossxNLOtwo{p_{123}}{p_3}{p_2}{p_1} =\left(\frac{S_\e}{16
\pi^2}\right)^2\, \frac{\left( -\sabc \right)^{-2\e}}
{\sab+\sac} \sum_{i=0}^4 \frac{f_{6.3,i}
\left(\frac{\sac}{\sabc},\frac{\sbc}{\sabc}\right)}{\e^i}\; + 
{\cal O}(\e) ,
\end{equation}
with:
\begin{eqnarray}
f_{6.3,4}(y,z) &=& 0\ , \\
f_{6.3,3}(y,z) &=& -\frac{1}{2} H(0;z)\ , \\
f_{6.3,2}(y,z) &=&  
          - H(0;y) H(0;z)
          - H(0;z) H(1-z;y)
          + H(0,0;z)
          - H(0,1;z)
          + \frac{\pi^2}{6}\ , \\
f_{6.3,1}(y,z) &=& 
          + 2 H(0;y) H(1,0;z)
          + 2 H(0;z) H(1-z,0;y)
          - 2 H(0;z) H(1-z,1-z;y) \nonumber \\
&&
          + 2 H(0,0;y) H(0;z)
          + 2 H(0,0;z) H(0;y)
          + 2 H(0,0;z) H(1-z;y)
          - 2 H(0,0,0;z)\nonumber \\
&&
          + 2 H(0,1,0;y)
          + 3 H(0,1,0;z)
          - 2 H(0,1,1;z)
          - 2 H(0,1-z;y) H(0;z)\nonumber \\
&&
          + 2 H(0,z;y) H(1;z)
          + 2 H(0,z,1-z;y)
          + 2 H(1;z) H(1-z,z;y)
          + 2 H(1,0;z) H(1-z;y)\nonumber \\
&&
          + 3 H(1,1,0;z)
          + 2 H(1-z,1,0;y)
          + 2 H(1-z,z,1-z;y)\nonumber \\
&&
          +\frac{\pi^2}{6} \Big[ 
                       + 2 H(0;y)
                       + 3 H(0;z)
                       + 3 H(1;z)
                       + 2 H(1-z;y) \Big] 
          + 5\zeta_3 \ ,\\
f_{6.3,0}(y,z) &=& 
          - 4 H(0;y) H(1,0,0;z)
          - 4 H(0;y) H(1,1,0;z)
          - 4 H(0;z) H(1-z,0,0;y)\nonumber \\
&&
          - 2 H(0;z) H(1-z,0,1-z;y)
          + 2 H(0;z) H(1-z,1,0;y)
          - 6 H(0;z) H(1-z,1-z,0;y)\nonumber \\
&&
          - 4 H(0;z) H(1-z,1-z,1-z;y)
          - 4 H(0,0;y) H(0,0;z)
          - 4 H(0,0;y) H(1,0;z)\nonumber \\
&&
          + 4 H(0,0;z) H(0,1-z;y)
          - 4 H(0,0;z) H(1-z,0;y)
          + 4 H(0,0;z) H(1-z,1-z;y)\nonumber \\
&&
          - 4 H(0,0,0;y) H(0;z)
          - 4 H(0,0,0;z) H(0;y)
          - 4 H(0,0,0;z) H(1-z;y)\nonumber \\
&&
          + 4 H(0,0,0,0;z)
          + 2 H(0,0,0,1;z)
          - 2 H(0,0,1;z) H(1-z;y)
          - 4 H(0,0,1,0;y)\nonumber \\
&&
          - 6 H(0,0,1,0;z)
          - 2 H(0,0,1-z;y) H(0;z)
          + 2 H(0,0,z;y) H(1;z)
          + 2 H(0,0,z,1-z;y)\nonumber \\
&&
          - 2 H(0,1;z) H(0,1-z;y)
          - 2 H(0,1;z) H(0,z;y)
          + 2 H(0,1;z) H(1-z,0;y)\nonumber \\
&&
          - 2 H(0,1;z) H(1-z,z;y)
          + 2 H(0,1,0;y) H(0;z)
          + 2 H(0,1,0;y) H(1;z)\nonumber \\
&&
          - 4 H(0,1,0;z) H(0;y)
          - 4 H(0,1,0;z) H(1-z;y)
          - 4 H(0,1,0,0;y)
          - 6 H(0,1,0,0;z)\nonumber \\
&&
          + 2 H(0,1,0,1;z)
          + 2 H(0,1,0,1-z;y)
          - 2 H(0,1,1,0;y)
          - 5 H(0,1,1,0;z)\nonumber \\
&&
          - 4 H(0,1,1,1;z)
          + 2 H(0,1,1-z,0;y)
          - 6 H(0,1-z;y) H(1,0;z)\nonumber \\
&&
          - 6 H(0,1-z,0;y) H(0;z)
          - 4 H(0,1-z,1,0;y)
          - 4 H(0,1-z,1-z;y) H(0;z)\nonumber \\
&&
          + 2 H(0,1-z,z;y) H(1;z)
          + 2 H(0,1-z,z,1-z;y)
          + 4 H(0,z;y) H(1,1;z)\nonumber \\
&&
          + 4 H(0,z,1-z;y) H(1;z)
          + 4 H(0,z,1-z,1-z;y)
          - 2 H(0,z,z;y) H(1;z)\nonumber \\
&&
          - 2 H(0,z,z,1-z;y)
          + 4 H(1;z) H(1-z,0,z;y)
          + 2 H(1;z) H(1-z,1,0;y)\nonumber \\
&&
          + 4 H(1;z) H(1-z,1-z,z;y)
          + 4 H(1;z) H(1-z,z,1-z;y)
          - 2 H(1;z) H(1-z,z,z;y)\nonumber \\
&&
          - 4 H(1,0;z) H(1-z,0;y)
          - 6 H(1,0;z) H(1-z,1-z;y)
          - 4 H(1,0,0;z) H(1-z;y)\nonumber \\
&&
          + 2 H(1,0,1;z) H(1-z;y)
          - 4 H(1,0,1,0;z)
          + 4 H(1,1;z) H(1-z,z;y)\nonumber \\
&&
          - 4 H(1,1,0;z) H(1-z;y)
          - 6 H(1,1,0,0;z)
          - 3 H(1,1,1,0;z)
          - 6 H(1-z,0,1,0;y)\nonumber \\
&&
          + 4 H(1-z,0,z,1-z;y)
          - 4 H(1-z,1,0,0;y)
          + 2 H(1-z,1,0,1-z;y)\nonumber \\
&&
          - 2 H(1-z,1,1,0;y)
          + 2 H(1-z,1,1-z,0;y)
          - 6 H(1-z,1-z,1,0;y)\nonumber \\
&&
          + 4 H(1-z,1-z,z,1-z;y)
          + 4 H(1-z,z,1-z,1-z;y)
          - 2 H(1-z,z,z,1-z;y)\nonumber \\
&&
          - \frac{\pi^4}{120}
          + \zeta_3 \Big[ 
              + 10 H(0;z)
              + H(1;z)
              + 6 H(1-z;y)
                    \Big]
          + \frac{\pi^2}{6} \Big[ 
              - 4 H(0;y) H(0;z) \nonumber \\
&&
              - 4 H(0;y) H(1;z)
              - 4 H(0;z) H(1-z;y)
              - 4 H(0,0;y)
              - 6 H(0,0;z)
              - 2 H(0,1;y)\nonumber \\
&&
              - 5 H(0,1;z)
              - 4 H(0,1-z;y)
              - 6 H(1;z) H(1-z;y)
              - 4 H(1,0;z)
              - 3 H(1,1;z)\nonumber \\
&&
              - 6 H(1-z,0;y)
              - 2 H(1-z,1;y)
              - 6 H(1-z,1-z;y) \Big]\ .
\end{eqnarray}

The two master integrals for the first 
seven-propagator crossed topology can be expressed 
in the following form, which makes the $p_1\leftrightarrow p_2$ interchange 
symmetry of (\ref{eq:sevena}) manifest:
\begin{eqnarray}
\lefteqn{\doublecrossNLO{p_{123}}{p_3}{p_2}{p_1} }\nonumber \\
&=&\left(\frac{S_\e}{16\pi^2}\right)^2\, \left( -\sabc \right)^{-2\e}
 \sum_{i=0}^4 \frac{1}{\e^i}\, 
\Bigg( \frac{\sabc}{\sab^2\sac\sbc} f_{7.3,i}\left(\frac{\sac}{\sabc},
                                                   \frac{\sbc}{\sabc}\right)
\left (1- 4\e + 16 \e^2 -64 \e^3\right)
\nonumber \\  
&&\hspace{4.5cm}
      + \frac{1}{\sab^2\sac}\left[f_{7.4,i}\left(\frac{\sac}{\sabc},
                                                   \frac{\sbc}{\sabc}\right)
\left (1- 4\e + 16 \e^2\right)
                                   +f_{7.5,i}\left(\frac{\sac}{\sabc},
                                                   \frac{\sbc}{\sabc}\right)
                              \right] \nonumber \\
&&\hspace{4.5cm}
      + \frac{1}{\sab^2\sbc}\left[f_{7.4,i}\left(\frac{\sbc}{\sabc},
                                                   \frac{\sac}{\sabc}\right)
\left (1- 4\e + 16 \e^2\right)
                                   +f_{7.5,i}\left(\frac{\sbc}{\sabc},
                                                   \frac{\sac}{\sabc}\right)
                              \right]\nonumber \\
&&\hspace{4.5cm}
      + \frac{1}{\sab^2\sabc} f_{6.2,i}\left(\frac{\sac}{\sabc},
                                                   \frac{\sbc}{\sabc}\right)
\Bigg) +
{\cal O}(\e) ,
\label{eq:npunit1}
\end{eqnarray}
and 
\begin{eqnarray}
\lefteqn{\doublecrossNLOtwo{p_{123}}{p_3}{p_2}{p_1} }\nonumber \\
&=&\left(\frac{S_\e}{16\pi^2}\right)^2\, \left( -\sabc \right)^{-2\e}
 \sum_{i=0}^4 \frac{1}{\e^i}\, 
\Bigg( \frac{\sabc}{\sab\sac\sbc} f_{7.3,i}\left(\frac{\sac}{\sabc},
                                                   \frac{\sbc}{\sabc}\right)
\left (1- 4\e + 16 \e^2 -64 \e^3\right)
\nonumber \\  
&&\hspace{4.5cm}
      + \frac{1}{\sab\sbc}\left[f_{7.4,i}\left(\frac{\sbc}{\sabc},
                                                   \frac{\sac}{\sabc}\right)
\left (1- 4\e + 16 \e^2\right)
                                   +f_{7.5,i}\left(\frac{\sbc}{\sabc},
                                                   \frac{\sac}{\sabc}\right)
                              \right] \nonumber \\
&&\hspace{4.5cm}
      + \frac{1}{\sab\sac} f_{7.4,i}\left(\frac{\sac}{\sabc},
                                                   \frac{\sbc}{\sabc}\right)
\left (1- 4\e + 16 \e^2\right)
                              \Bigg) +
{\cal O}(\e) ,
\end{eqnarray}
with
\begin{eqnarray}
f_{7.3,4}(y,z) &=& 0 \ ,\\
f_{7.3,3}(y,z) &=& -\frac{3}{2}\ , \\
f_{7.3,2}(y,z) &=& 3 H(0;y) + 3 H(0;z) \\
f_{7.3,1}(y,z) &=&           
          - 6 H(0;y) H(0;z)
          - 6 H(0,0;y)
          - 6 H(0,0;z)
          - 6 H(1,0;y)
          - 6 H(1,0;z) -\pi^2\ , \\
f_{7.3,0}(y,z) &=&  
          + 12 H(0;y) H(1,0,z)
          + 12 H(0;z) H(1-z,0,y)
          + 12 H(0,0;y) H(0,z) \nonumber \\
&&
          + 12 H(0,0;z) H(0,y)
          + 12 H(0,0,0;y)
          + 12 H(0,0,0;z)
          + 12 H(0,1,0;y)
          + 12 H(0,1,0;z)\nonumber \\
&&
          + 12 H(1,0;z) H(1-z;y)
          + 12 H(1,0,0;y)
          + 12 H(1,0,0;z)
          + 12 H(1,1,0;z)\nonumber \\
&&
          + 12 H(1-z,1,0;y)
          + 9 \zeta_3 
          + 2\pi^2 \left[           
              + H(0;y)
              + H(0;z)
              + H(1;z)
              + H(1-z;y) \right] \ ,\\
f_{7.4,4}(y,z) &=& 0 \ , \\
f_{7.4,3}(y,z) &=& 0 \ , \\
f_{7.4,2}(y,z) &=& -3 H(0;z) - 3 H(1;z) - 3 H(1-z;y) \ , \\ 
f_{7.4,1}(y,z) &=&           
          + 6 H(0;y) H(0;z)
          + 6 H(0;y) H(1;z)
          + 6 H(0,0;z)
          - 6 H(0,1;z)
          + 6 H(0,1-z;y)\nonumber \\
&&
          - 6 H(1;z) H(1-z;y)
          - 6 H(1;z) H(z;y)
          + 6 H(1,0;z)
          - 6 H(1,1;z)
          + 6 H(1-z,0;y)\nonumber \\
&&
          - 6 H(1-z,1-z;y)
          - 6 H(z,1-z;y) + \pi^2 \ , \\
f_{7.4,0}(y,z) &=&           
          - 12 H(0;y) H(1,0;z)
          + 12 H(0;y) H(1,1;z)
          - 12 H(0;z) H(1-z,0;y)\nonumber \\
&&
          - 12 H(0,0;y) H(0;z)
          - 12 H(0,0;y) H(1;z)
          - 12 H(0,0;z) H(0;y)
          - 12 H(0,0,0;z)\nonumber \\
&&
          - 12 H(0,0,1;z)
          - 12 H(0,0,1-z;y)
          + 12 H(0,1;z) H(0;y)
          - 12 H(0,1;z) H(1-z;y)\nonumber \\
&&
          - 12 H(0,1,0;z)
          - 12 H(0,1,1;z)
          + 12 H(0,1-z;y) H(1;z)
          - 12 H(0,1-z,0;y)\nonumber \\
&&
          + 12 H(0,1-z,1-z;y)
          + 12 H(0,z;y) H(1;z)
          + 12 H(0,z,1-z;y)\nonumber \\
&&
          + 12 H(1;z) H(1-z,0;y)
          - 12 H(1;z) H(1-z,1-z;y)
          - 12 H(1;z) H(1-z,z;y)\nonumber \\
&&
          - 12 H(1;z) H(z,1-z;y)
          - 12 H(1,0;z) H(1-z;y)
          - 12 H(1,0,0;z)
          - 12 H(1,0,1;z)\nonumber \\
&&
          - 12 H(1,1;z) H(1-z;y)
          - 12 H(1,1;z) H(z;y)
          - 12 H(1,1,0;z)
          - 12 H(1,1,1;z)\nonumber \\
&&
          - 12 H(1-z,0,0;y)
          + 12 H(1-z,0,1-z;y)
          - 24 H(1-z,1,0;y)
          + 12 H(1-z,1-z,0;y)\nonumber \\
&&
          - 12 H(1-z,1-z,1-z;y)
          - 12 H(1-z,z,1-z;y)
          - 12 H(z,1-z,1-z;y)\nonumber \\
&&
          - 2\pi^2 \left[
             + H(0;y)
             + H(0;z)
             + H(1;z)
             + H(1-z;y)
                   \right] \ , \\
f_{7.5,4}(y,z) &=& -\frac{1}{8}\ ,  \\
f_{7.5,3}(y,z) &=& 
          + \frac{1}{4} H(0;y)
          - \frac{3}{4} H(0;z)
          - \frac{1}{4} H(1;z)
          - \frac{1}{4} H(1-z;y)   \ ,      \\
f_{7.5,2}(y,z) &=& 
          + \frac{3}{2} H(0;y) H(0;z)
          + \frac{1}{2} H(0;y) H(1;z)
          - \frac{3}{2} H(0;z) H(1-z;y)
          - \frac{1}{2} H(0,0;y)
          + \frac{3}{2} H(0,0;z)\nonumber \\
&&
          + \frac{3}{2} H(0,1;z)
          + \frac{1}{2} H(0,1-z;y)
          - \frac{1}{2} H(1;z) H(1-z;y)
          + 3 H(1;z) H(z;y)
          - 2 H(1,0;y)\nonumber \\
&&
          - \frac{3}{2} H(1,0;z)
          - \frac{1}{2} H(1,1;z)
          + \frac{1}{2} H(1-z,0;y)
          - \frac{1}{2} H(1-z,1-z;y)
          + 3 H(z,1-z;y)\nonumber \\
&&        + \frac{\pi^2}{3} \ , \\
f_{7.5,1}(y,z) &=& 
          + 3 H(0;y) H(1,0;z)
          + H(0;y) H(1,1;z)
          - 4 H(0;z) H(1,0;y)
          + 3 H(0;z) H(1-z,0;y)\nonumber \\
&&
          - 3 H(0;z) H(1-z,1-z;y)
          - 6 H(0;z) H(z,1-z;y)
          - 3 H(0,0;y) H(0;z)\nonumber \\
&&
          - H(0,0;y) H(1;z)
          - 3 H(0,0;z) H(0;y)
          + 3 H(0,0;z) H(1-z;y)
          + H(0,0,0;y)\nonumber \\
&&
          - 3 H(0,0,0;z)
          + 6 H(0,0,1;z)
          - H(0,0,1-z;y)
          - 3 H(0,1;z) H(0;y)\nonumber \\
&&
          + 3 H(0,1;z) H(1-z;y)
          + 9 H(0,1;z) H(z;y)
          + H(0,1,0;y)
          - 9 H(0,1,0;z)
          + 3 H(0,1,1;z)\nonumber \\
&&
          + 3 H(0,1-z;y) H(0;z)
          + H(0,1-z;y) H(1;z)
          - H(0,1-z,0;y)
          + H(0,1-z,1-z;y)\nonumber \\
&&
          - 6 H(0,z;y) H(1;z)
          - 6 H(0,z,1-z;y)
          + H(1;z) H(1-z,0;y)\nonumber \\
&&
          - H(1;z) H(1-z,1-z;y)
          + 6 H(1;z) H(1-z,z;y)
          - 6 H(1;z) H(z,0;y)\nonumber \\
&&
          + 6 H(1;z) H(z,1-z;y)
          + 15 H(1;z) H(z,z;y)
          - 4 H(1,0;y) H(1;z)\nonumber \\
&&
          - 3 H(1,0;z) H(1-z;y)
          - 6 H(1,0;z) H(z;y)
          + 4 H(1,0,0;y)
          + 3 H(1,0,0;z)\nonumber \\
&&
          + 3 H(1,0,1;z)
          - 4 H(1,0,1-z;y)
          - H(1,1;z) H(1-z;y)
          + 6 H(1,1;z) H(z;y)\nonumber \\
&&
          + 3 H(1,1,0;y)
          - 3 H(1,1,0;z)
          - H(1,1,1;z)
          - 4 H(1,1-z,0;y)
          - H(1-z,0,0;y)\nonumber \\
&&
          + H(1-z,0,1-z;y)
          - 4 H(1-z,1,0;y)
          + H(1-z,1-z,0;y)
          - H(1-z,1-z,1-z;y)\nonumber \\
&&
          + 6 H(1-z,z,1-z;y)
          - 6 H(z,0,1-z;y)
          - 6 H(z,1-z,0;y)
          + 6 H(z,1-z,1-z;y)\nonumber \\
&&
          + 15 H(z,z,1-z;y)
          + \frac{17}{4} \zeta_3 
          + \frac{\pi^2}{6} \big[ 
            - 4 H(0;y)
            - 3 H(0;z)
            + 3 H(1;y)
            + 4 H(1;z)\nonumber \\
&&
            + 4 H(1-z;y)      \big]\ ,        \\
f_{7.5,0}(y,z) &=& 
          - 6 H(0;y) H(1,0,0;z)
          - 6 H(0;y) H(1,0,1;z)
          + 6 H(0;y) H(1,1,0;z)\nonumber \\
&&
          + 2 H(0;y) H(1,1,1;z)
          + 8 H(0;z) H(1,0,0;y)
          - 2 H(0;z) H(1,0,1-z;y)\nonumber \\
&&
          + 8 H(0;z) H(1,1-z,0;y)
          - 6 H(0;z) H(1-z,0,0;y)
          + 6 H(0;z) H(1-z,0,1-z;y)\nonumber \\
&&
          - 8 H(0;z) H(1-z,1,0;y)
          + 6 H(0;z) H(1-z,1-z,0;y)\nonumber \\
&&
          - 6 H(0;z) H(1-z,1-z,1-z;y)
          - 12 H(0;z) H(1-z,z,1-z;y)\nonumber \\
&&
          - 6 H(0;z) H(z,0,1-z;y)
          + 8 H(0;z) H(z,1-z,0;y)
          - 12 H(0;z) H(z,1-z,1-z;y)\nonumber \\
&&
          - 6 H(0;z) H(z,z,1-z;y)
          + 6 H(0,0;y) H(0,0;z)
          + 6 H(0,0;y) H(0,1;z)\nonumber \\
&&
          - 6 H(0,0;y) H(1,0;z)
          - 2 H(0,0;y) H(1,1;z)
          - 6 H(0,0;z) H(0,1-z;y)\nonumber \\
&&
          + 8 H(0,0;z) H(1,0;y)
          - 6 H(0,0;z) H(1-z,0;y)
          + 6 H(0,0;z) H(1-z,1-z;y)\nonumber \\
&&
          + 12 H(0,0;z) H(z,1-z;y)
          + 6 H(0,0,0;y) H(0;z)
          + 2 H(0,0,0;y) H(1;z)\nonumber \\
&&
          + 6 H(0,0,0;z) H(0;y)
          - 6 H(0,0,0;z) H(1-z;y)
          - 2 H(0,0,0,0;y)\nonumber \\
&&
          + 6 H(0,0,0,0;z)
          + 15 H(0,0,0,1;z)
          + 2 H(0,0,0,1-z;y)
          - 12 H(0,0,1;z) H(0;y)\nonumber \\
&&
          + 6 H(0,0,1;z) H(1;y)
          + 12 H(0,0,1;z) H(1-z;y)
          + 9 H(0,0,1;z) H(z;y)\nonumber \\
&&
          + 4 H(0,0,1,0;y)
          - 6 H(0,0,1,0;z)
          + 12 H(0,0,1,1;z)
          - 6 H(0,0,1-z;y) H(0;z)\nonumber \\
&&
          - 2 H(0,0,1-z;y) H(1;z)
          + 2 H(0,0,1-z,0;y)
          - 2 H(0,0,1-z,1-z;y)\nonumber \\
&&
          + 12 H(0,0,z;y) H(1;z)
          + 12 H(0,0,z,1-z;y)
          - 6 H(0,1;z) H(0,1-z;y)\nonumber \\
&&
          - 12 H(0,1;z) H(0,z;y)
          - 8 H(0,1;z) H(1,0;y)
          + 6 H(0,1;z) H(1,1-z;y)\nonumber \\
&&
          - 6 H(0,1;z) H(1-z,0;y)
          + 6 H(0,1;z) H(1-z,1-z;y)
          + 18 H(0,1;z) H(1-z,z;y)\nonumber \\
&&
          - 12 H(0,1;z) H(z,0;y)
          + 12 H(0,1;z) H(z,1-z;y)
          + 3 H(0,1;z) H(z,z;y)\nonumber \\
&&
          + 8 H(0,1,0;y) H(0;z)
          + 8 H(0,1,0;y) H(1;z)
          + 18 H(0,1,0;z) H(0;y)\nonumber \\
&&
          + 10 H(0,1,0;z) H(1;y)
          - 18 H(0,1,0;z) H(1-z;y)
          - 8 H(0,1,0;z) H(z;y)\nonumber \\
&&
          - 2 H(0,1,0,0;y)
          + 18 H(0,1,0,0;z)
          + 12 H(0,1,0,1;z)
          + 8 H(0,1,0,1-z;y)\nonumber \\
&&
          - 6 H(0,1,1;z) H(0;y)
          + 6 H(0,1,1;z) H(1-z;y)
          + 18 H(0,1,1;z) H(z;y)\nonumber \\
&&
          - 3 H(0,1,1,0;y)
          + 6 H(0,1,1,0;z)
          + 6 H(0,1,1,1;z)
          + 8 H(0,1,1-z,0;y)\nonumber \\
&&
          + 6 H(0,1-z;y) H(1,0;z)
          + 2 H(0,1-z;y) H(1,1;z)
          - 6 H(0,1-z,0;y) H(0;z)\nonumber \\
&&
          - 2 H(0,1-z,0;y) H(1;z)
          + 2 H(0,1-z,0,0;y)
          - 2 H(0,1-z,0,1-z;y)\nonumber \\
&&
          + 2 H(0,1-z,1,0;y)
          + 6 H(0,1-z,1-z;y) H(0;z)
          + 2 H(0,1-z,1-z;y) H(1;z)\nonumber \\
&&
          - 2 H(0,1-z,1-z,0;y)
          + 2 H(0,1-z,1-z,1-z;y)
          - 12 H(0,1-z,z;y) H(1;z)\nonumber \\
&&
          - 12 H(0,1-z,z,1-z;y)
          + 12 H(0,z;y) H(1,0;z)
          - 12 H(0,z;y) H(1,1;z)\nonumber \\
&&
          + 6 H(0,z,0;y) H(1;z)
          + 6 H(0,z,0,1-z;y)
          + 12 H(0,z,1-z;y) H(0;z)\nonumber \\
&&
          - 12 H(0,z,1-z;y) H(1;z)
          + 6 H(0,z,1-z,0;y)
          - 12 H(0,z,1-z,1-z;y)\nonumber \\
&&
          - 24 H(0,z,z;y) H(1;z)
          - 24 H(0,z,z,1-z;y)
          - 2 H(1;z) H(1-z,0,0;y)\nonumber \\
&&
          + 2 H(1;z) H(1-z,0,1-z;y)
          - 12 H(1;z) H(1-z,0,z;y)
          - 8 H(1;z) H(1-z,1,0;y)\nonumber \\
&&
          + 2 H(1;z) H(1-z,1-z,0;y)
          - 2 H(1;z) H(1-z,1-z,1-z;y)\nonumber \\
&&
          + 12 H(1;z) H(1-z,1-z,z;y)
          - 12 H(1;z) H(1-z,z,0;y)\nonumber \\
&&
          + 12 H(1;z) H(1-z,z,1-z;y)
          + 30 H(1;z) H(1-z,z,z;y)
          + 12 H(1;z) H(z,0,0;y)\nonumber \\
&&
          - 12 H(1;z) H(z,0,1-z;y)
          - 6 H(1;z) H(z,0,z;y)
          - 12 H(1;z) H(z,1-z,0;y)\nonumber \\
&&
          + 12 H(1;z) H(z,1-z,1-z;y)
          + 24 H(1;z) H(z,1-z,z;y)
          - 6 H(1;z) H(z,z,0;y)\nonumber \\
&&
          + 30 H(1;z) H(z,z,1-z;y)
          + 9 H(1;z) H(z,z,z;y)
          + 8 H(1,0;y) H(1,0;z)\nonumber \\
&&
          - 8 H(1,0;y) H(1,1;z)
          + 10 H(1,0;z) H(1,1-z;y)
          + 6 H(1,0;z) H(1-z,0;y)\nonumber \\
&&
          - 6 H(1,0;z) H(1-z,1-z;y)
          - 12 H(1,0;z) H(1-z,z;y)
          + 8 H(1,0;z) H(z,0;y)\nonumber \\
&&
          + 2 H(1,0;z) H(z,1-z;y)
          - 6 H(1,0;z) H(z,z;y)
          + 8 H(1,0,0;y) H(1;z)\nonumber \\
&&
          + 6 H(1,0,0;z) H(1-z;y)
          + 12 H(1,0,0;z) H(z;y)
          - 8 H(1,0,0,0;y)\nonumber \\
&&
          - 6 H(1,0,0,0;z)
          + 12 H(1,0,0,1;z)
          + 8 H(1,0,0,1-z;y)
          + 6 H(1,0,1;z) H(1;y)\nonumber \\
&&
          + 6 H(1,0,1;z) H(1-z;y)
          + 12 H(1,0,1;z) H(z;y)
          + 4 H(1,0,1,0;y)\nonumber \\
&&
          - 18 H(1,0,1,0;z)
          + 6 H(1,0,1,1;z)
          - 8 H(1,0,1-z;y) H(1;z)
          + 8 H(1,0,1-z,0;y)\nonumber \\
&&
          - 8 H(1,0,1-z,1-z;y)
          - 6 H(1,0,z;y) H(1;z)
          - 6 H(1,0,z,1-z;y)\nonumber \\
&&
          + 2 H(1,1;z) H(1-z,0;y)
          - 2 H(1,1;z) H(1-z,1-z;y)
          + 12 H(1,1;z) H(1-z,z;y)\nonumber \\
&&
          - 12 H(1,1;z) H(z,0;y)
          + 12 H(1,1;z) H(z,1-z;y)
          + 30 H(1,1;z) H(z,z;y)\nonumber \\
&&
          + 10 H(1,1,0;z) H(1;y)
          - 6 H(1,1,0;z) H(1-z;y)
          + 2 H(1,1,0;z) H(z;y)\nonumber \\
&&
          - 6 H(1,1,0,0;y)
          + 6 H(1,1,0,0;z)
          + 6 H(1,1,0,1;z)
          - 2 H(1,1,1;z) H(1-z;y)\nonumber \\
&&
          + 12 H(1,1,1;z) H(z;y)
          + 3 H(1,1,1,0;y)
          - 6 H(1,1,1,0;z)
          - 2 H(1,1,1,1;z)\nonumber \\
&&
          - 8 H(1,1-z,0;y) H(1;z)
          + 8 H(1,1-z,0,0;y)
          - 8 H(1,1-z,0,1-z;y)\nonumber \\
&&
          + 16 H(1,1-z,1,0;y)
          - 8 H(1,1-z,1-z,0;y)
          + 6 H(1,1-z,z;y) H(1;z)\nonumber \\
&&
          + 6 H(1,1-z,z,1-z;y)
          + 2 H(1-z,0,0,0;y)
          - 2 H(1-z,0,0,1-z;y)\nonumber \\
&&
          + 2 H(1-z,0,1,0;y)
          - 2 H(1-z,0,1-z,0;y)
          + 2 H(1-z,0,1-z,1-z;y)\nonumber \\
&&
          - 12 H(1-z,0,z,1-z;y)
          + 8 H(1-z,1,0,0;y)
          - 8 H(1-z,1,0,1-z;y)\nonumber \\
&&
          + 6 H(1-z,1,1,0;y)
          - 8 H(1-z,1,1-z,0;y)
          - 2 H(1-z,1-z,0,0;y)\nonumber \\
&&
          + 2 H(1-z,1-z,0,1-z;y)
          - 8 H(1-z,1-z,1,0;y)
          + 2 H(1-z,1-z,1-z,0;y)\nonumber \\
&&
          - 2 H(1-z,1-z,1-z,1-z;y)
          + 12 H(1-z,1-z,z,1-z;y)\nonumber \\
&&
          - 12 H(1-z,z,0,1-z;y)
          - 12 H(1-z,z,1-z,0;y)
          + 12 H(1-z,z,1-z,1-z;y)\nonumber \\
&&
          + 30 H(1-z,z,z,1-z;y)
          + 12 H(z,0,0,1-z;y)
          - 14 H(z,0,1,0;y)\nonumber \\
&&
          + 12 H(z,0,1-z,0;y)
          - 12 H(z,0,1-z,1-z;y)
          - 6 H(z,0,z,1-z;y)\nonumber \\
&&
          + 12 H(z,1-z,0,0;y)
          - 12 H(z,1-z,0,1-z;y)
          + 14 H(z,1-z,1,0;y)\nonumber \\
&&
          - 12 H(z,1-z,1-z,0;y)
          + 12 H(z,1-z,1-z,1-z;y)
          + 24 H(z,1-z,z,1-z;y)\nonumber \\
&&
          - 6 H(z,z,0,1-z;y)
          - 6 H(z,z,1-z,0;y)
          + 30 H(z,z,1-z,1-z;y)\nonumber \\
&&
          + 9 H(z,z,z,1-z;y)
          + \frac{71\pi^4}{240}
          + \frac{\zeta_3}{2} \big[ 
                 - 17 H(0;y)
                 + 9 H(0;z)
                 + 26 H(1;y)
                 + 17 H(1;z)\nonumber \\
&&
                 + 17 H(1-z;y)
                              \big]
          + \frac{\pi^2}{6} \big[
          + 6 H(0;y) H(0;z)
          - 8 H(0;y) H(1;z)
          + 10 H(0;z) H(1;y)\nonumber \\
&&
          - 6 H(0;z) H(1-z;y)
          + 8 H(0,0;y)
          + 6 H(0,0;z)
          - 3 H(0,1;y)
          + 18 H(0,1;z)\nonumber \\
&&
          - 8 H(0,1-z;y)
          + 10 H(1;y) H(1;z)
          + 8 H(1;z) H(1-z;y)
          + 20 H(1;z) H(z;y)\nonumber \\
&&
          + 12 H(1,0;y)
          - 6 H(1,0;z)
          + 3 H(1,1;y)
          + 8 H(1,1;z)
          + 10 H(1,1-z;y)\nonumber \\
&&
          - 8 H(1-z,0;y)
          + 6 H(1-z,1;y)
          + 8 H(1-z,1-z;y)
          + 20 H(z,1-z;y)
                            \big]\ .
\end{eqnarray}

The two master integrals for the second seven-propagator 
crossed topology are expressed as follows. Again, 
the $p_1\leftrightarrow p_2$ interchange 
symmetry of (\ref{eq:sevenc}) becomes apparent:
\begin{eqnarray}
\lefteqn{\doublecrossxNLO{p_{123}}{p_3}{p_2}{p_1} }\nonumber \\
&=&\left(\frac{S_\e}{16\pi^2}\right)^2\, \left( -\sabc \right)^{-2\e}
 \sum_{i=0}^4 \frac{1}{\e^i}\, 
\Bigg( \frac{1}{\sac\sbc\sabc} f_{7.6,i}\left(\frac{\sac}{\sabc},
                                                   \frac{\sbc}{\sabc}\right)
\nonumber \\  
&&\hspace{4.5cm}
      + \frac{1}{\sab\sac\sabc} f_{7.7,i}\left(\frac{\sac}{\sabc},
                                                   \frac{\sbc}{\sabc}\right)
 \nonumber \\
&&\hspace{4.5cm}
      + \frac{1}{\sab\sbc\sabc} f_{7.7,i}\left(\frac{\sbc}{\sabc},
                                                   \frac{\sac}{\sabc}\right)
 \nonumber \\
&&\hspace{4.5cm}
      + \frac{1}{\sac(\sac+\sbc)\sabc} f_{7.8,i}\left(\frac{\sac}{\sabc},
                                                   \frac{\sbc}{\sabc}\right)
 \nonumber \\
&&\hspace{4.5cm}
      + \frac{1}{\sbc(\sac+\sbc)\sabc} f_{7.8,i}\left(\frac{\sbc}{\sabc},
                                                   \frac{\sac}{\sabc}\right)
\Bigg) +
{\cal O}(\e) ,
\label{eq:npunit2}
\end{eqnarray}
and 
\begin{eqnarray}
\lefteqn{\doublecrossxNLOtwo{p_{123}}{p_3}{p_2}{p_1} }\nonumber \\
&=&\left(\frac{S_\e}{16\pi^2}\right)^2\, \left( -\sabc \right)^{-2\e}
 \sum_{i=0}^4 \frac{1}{\e^i}\, 
\Bigg( \frac{1}{\sab\sac} f_{7.7,i}\left(\frac{\sac}{\sabc},
                                                   \frac{\sbc}{\sabc}\right)
\nonumber \\  
&&\hspace{4.5cm}
      - \frac{1}{\sac\sabc} f_{6.2,i}\left(\frac{\sab}{\sabc},
                                                   \frac{\sbc}{\sabc}\right)
                             \Bigg) +
{\cal O}(\e) ,
\end{eqnarray}
with
\begin{eqnarray}
f_{7.6,4}(y,z) &=& -1 \ ,\\
f_{7.6,3}(y,z) &=& 2 H(0;y) + 2 H(0;z)\ , \\
f_{7.6,2}(y,z) &=&           
          - 4 H(0;y) H(0;z)
          - 4 H(0,0;y)
          - 4 H(0,0;z)
          + 3 H(1;z) H(1-z;y)
          - 3 H(1,0;y)\nonumber \\
&&
          - 3 H(1,0;z)
          + 3 H(1,1;z)
          + 3 H(1-z,1-z;y)
          - \frac{\pi^2}{3} \ ,\\
f_{7.6,1}(y,z) &=&           
          + 6 H(0;y) H(1,0;z)
          - 4 H(0;y) H(1,1;z)
          + 4 H(0;z) H(1,0;y)
          + 2 H(0;z) H(1-z,0;y)\nonumber \\
&&
          - 4 H(0;z) H(1-z,1-z;y)
          + 8 H(0,0;y) H(0;z)
          + 8 H(0,0;z) H(0;y)
          + 8 H(0,0,0;y)\nonumber \\
&&
          + 8 H(0,0,0;z)
          + 6 H(0,1;z) H(1-z;y)
          + 6 H(0,1,0;y)
          + 6 H(0,1,0;z)\nonumber \\
&&
          - 2 H(0,1,1;z)
          - 4 H(0,1-z;y) H(1;z)
          - 4 H(0,1-z,1-z;y)
          - 4 H(1;z) H(1-z,0;y)\nonumber \\
&&
          + 9 H(1;z) H(1-z,1-z;y)
          + 10 H(1;z) H(1-z,z;y)
          + 2 H(1;z) H(z,1-z;y)\nonumber \\
&&
          + 4 H(1,0;y) H(1;z)
          + 2 H(1,0;z) H(1-z;y)
          + 6 H(1,0,0;y)
          + 6 H(1,0,0;z)\nonumber \\
&&
          + 10 H(1,0,1;z)
          + 4 H(1,0,1-z;y)
          + 9 H(1,1;z) H(1-z;y)
          + 2 H(1,1;z) H(z;y)\nonumber \\
&&
          + H(1,1,0;y)
          + 7 H(1,1,0;z)
          + 9 H(1,1,1;z)
          + 4 H(1,1-z,0;y)\nonumber \\
&&
          - 4 H(1-z,0,1-z;y)
          + 6 H(1-z,1,0;y)
          - 4 H(1-z,1-z,0;y)\nonumber \\
&&
          + 9 H(1-z,1-z,1-z;y)
          + 10 H(1-z,z,1-z;y)
          + 2 H(z,1-z,1-z;y)\nonumber \\
&&
          + 16 \zeta_3 
          +\frac{\pi^2}{6} \big[
            + 4 H(0;y)
            + 4 H(0;z)
            + H(1;y)
            + 3 H(1;z)
            + 2 H(1-z;y)
                           \big] \ ,\\
f_{7.6,0}(y,z) &=&           
          - 12 H(0;y) H(1,0,0;z)
          - 16 H(0;y) H(1,0,1;z)
          - 14 H(0;y) H(1,1,0;z)\nonumber \\
&&
          - 12 H(0;y) H(1,1,1;z)
          - 8 H(0;z) H(1,0,0;y)
          - 2 H(0;z) H(1,0,1-z;y)\nonumber \\
&&
          - 8 H(0;z) H(1,1-z,0;y)
          - 4 H(0;z) H(1-z,0,0;y)
          - 14 H(0;z) H(1-z,0,1-z;y)\nonumber \\
&&
          - 4 H(0;z) H(1-z,1,0;y)
          - 2 H(0;z) H(1-z,1-z,0;y)\nonumber \\
&&
          - 12 H(0;z) H(1-z,1-z,1-z;y)
          - 16 H(0,0;y) H(0,0;z)
          - 12 H(0,0;y) H(1,0;z)\nonumber \\
&&
          + 8 H(0,0;y) H(1,1;z)
          - 8 H(0,0;z) H(1,0;y)
          - 4 H(0,0;z) H(1-z,0;y)\nonumber \\
&&
          + 8 H(0,0;z) H(1-z,1-z;y)
          - 16 H(0,0,0;y) H(0;z)
          - 16 H(0,0,0;z) H(0;y)\nonumber \\
&&
          - 16 H(0,0,0,0;y)
          - 16 H(0,0,0,0;z)
          - 12 H(0,0,1;z) H(1;y)
          + 14 H(0,0,1;z) H(1-z;y)\nonumber \\
&&
          - 12 H(0,0,1,0;y)
          - 12 H(0,0,1,0;z)
          + 4 H(0,0,1,1;z)
          + 8 H(0,0,1-z;y) H(1;z)\nonumber \\
&&
          + 8 H(0,0,1-z,1-z;y)
          - 12 H(0,1;z) H(0,1-z;y)
          + 10 H(0,1;z) H(1,0;y)\nonumber \\
&&
          - 12 H(0,1;z) H(1,1-z;y)
          - 14 H(0,1;z) H(1-z,0;y)
          + 13 H(0,1;z) H(1-z,1-z;y)\nonumber \\
&&
          + 6 H(0,1;z) H(1-z,z;y)
          + 4 H(0,1;z) H(z,1-z;y)
          - 4 H(0,1,0;y) H(0;z)\nonumber \\
&&
          - 4 H(0,1,0;y) H(1;z)
          - 12 H(0,1,0;z) H(0;y)
          - 6 H(0,1,0;z) H(1;y)\nonumber \\
&&
          + 2 H(0,1,0;z) H(1-z;y)
          - 12 H(0,1,0,0;y)
          - 12 H(0,1,0,0;z)
          - 4 H(0,1,0,1;z)\nonumber \\
&&
          - 4 H(0,1,0,1-z;y)
          + 11 H(0,1,1;z) H(1-z;y)
          - 4 H(0,1,1;z) H(z;y)\nonumber \\
&&
          + 2 H(0,1,1,0;y)
          - 6 H(0,1,1,0;z)
          - 6 H(0,1,1,1;z)
          - 4 H(0,1,1-z,0;y)\nonumber \\
&&
          - 12 H(0,1-z;y) H(1,0;z)
          - 12 H(0,1-z;y) H(1,1;z)
          - 8 H(0,1-z,0;y) H(0;z)\nonumber \\
&&
          + 8 H(0,1-z,0;y) H(1;z)
          + 8 H(0,1-z,0,1-z;y)
          - 12 H(0,1-z,1,0;y)\nonumber \\
&&
          - 12 H(0,1-z,1-z;y) H(1;z)
          + 8 H(0,1-z,1-z,0;y)
          - 12 H(0,1-z,1-z,1-z;y)\nonumber \\
&&
          - 12 H(0,1-z,z;y) H(1;z)
          - 12 H(0,1-z,z,1-z;y)
          + 8 H(1;z) H(1-z,0,0;y)\nonumber \\
&&
          - 12 H(1;z) H(1-z,0,1-z;y)
          - 4 H(1;z) H(1-z,1,0;y)
          - 12 H(1;z) H(1-z,1-z,0;y)\nonumber \\
&&
          + 21 H(1;z) H(1-z,1-z,1-z;y)
          + 25 H(1;z) H(1-z,1-z,z;y)\nonumber \\
&&
          + 23 H(1;z) H(1-z,z,1-z;y)
          + 6 H(1;z) H(1-z,z,z;y)
          + 6 H(1;z) H(z,1-z,1-z;y)\nonumber \\
&&
          + 4 H(1;z) H(z,1-z,z;y)
          - 4 H(1;z) H(z,z,1-z;y)
          - 8 H(1,0;y) H(1,0;z)\nonumber \\
&&
          + 10 H(1,0;y) H(1,1;z)
          - 6 H(1,0;z) H(1,1-z;y)
          - 6 H(1,0;z) H(1-z,0;y)\nonumber \\
&&
          - 4 H(1,0;z) H(1-z,1-z;y)
          - 8 H(1,0,0;y) H(1;z)
          - 4 H(1,0,0;z) H(1-z;y)\nonumber \\
&&
          - 12 H(1,0,0,0;y)
          - 12 H(1,0,0,0;z)
          + 18 H(1,0,0,1;z)
          - 8 H(1,0,0,1-z;y)\nonumber \\
&&
          - 12 H(1,0,1;z) H(1;y)
          + 9 H(1,0,1;z) H(1-z;y)
          + 4 H(1,0,1;z) H(z;y)\nonumber \\
&&
          + 2 H(1,0,1,0;y)
          - 4 H(1,0,1,0;z)
          + 21 H(1,0,1,1;z)
          + 10 H(1,0,1-z;y) H(1;z)\nonumber \\
&&
          - 8 H(1,0,1-z,0;y)
          + 10 H(1,0,1-z,1-z;y)
          + 12 H(1,0,z;y) H(1;z)\nonumber \\
&&
          + 12 H(1,0,z,1-z;y)
          - 12 H(1,1;z) H(1-z,0;y)
          + 21 H(1,1;z) H(1-z,1-z;y)\nonumber \\
&&
          + 23 H(1,1;z) H(1-z,z;y)
          + 6 H(1,1;z) H(z,1-z;y)
          - 4 H(1,1;z) H(z,z;y)\nonumber \\
&&
          - 6 H(1,1,0;z) H(1;y)
          - 6 H(1,1,0;z) H(1-z;y)
          - 2 H(1,1,0,0;y)
          - 14 H(1,1,0,0;z)\nonumber \\
&&
          + 7 H(1,1,0,1;z)
          + 21 H(1,1,1;z) H(1-z;y)
          + 6 H(1,1,1;z) H(z;y)
          - H(1,1,1,0;y)\nonumber \\
&&
          - 15 H(1,1,1,0;z)
          + 21 H(1,1,1,1;z)
          + 10 H(1,1-z,0;y) H(1;z)
          - 8 H(1,1-z,0,0;y)\nonumber \\
&&
          + 10 H(1,1-z,0,1-z;y)
          - 18 H(1,1-z,1,0;y)
          + 10 H(1,1-z,1-z,0;y)\nonumber \\
&&
          - 12 H(1,1-z,z;y) H(1;z)
          - 12 H(1,1-z,z,1-z;y)
          + 8 H(1-z,0,0,1-z;y)\nonumber \\
&&
          - 6 H(1-z,0,1,0;y)
          + 8 H(1-z,0,1-z,0;y)
          - 12 H(1-z,0,1-z,1-z;y)\nonumber \\
&&
          - 12 H(1-z,1,0,0;y)
          - 4 H(1-z,1,0,1-z;y)
          + 2 H(1-z,1,1,0;y)\nonumber \\
&&
          - 4 H(1-z,1,1-z,0;y)
          + 8 H(1-z,1-z,0,0;y)
          - 12 H(1-z,1-z,0,1-z;y)\nonumber \\
&&
          + 8 H(1-z,1-z,1,0;y)
          - 12 H(1-z,1-z,1-z,0;y)\nonumber \\
&&
          + 21 H(1-z,1-z,1-z,1-z;y)
          + 25 H(1-z,1-z,z,1-z;y)\nonumber \\
&&
          + 23 H(1-z,z,1-z,1-z;y)
          + 6 H(1-z,z,z,1-z;y)\nonumber \\
&&
          + 6 H(z,1-z,1-z,1-z;y)
          + 4 H(z,1-z,z,1-z;y)
          - 4 H(z,z,1-z,1-z;y)\nonumber \\
&&
          + \frac{\mbox{2}\pi^4}{5}
          + \zeta_3 \big[ 
            - 32 H(0;y)
            - 32 H(0;z)
            - 11 H(1;y)
            - 15 H(1;z)
            - 4 H(1-z;y)
                    \big]\nonumber \\
&&
          + \frac{\pi^2}{6} \big[
            - 8 H(0;y) H(0;z)
            - 6 H(0;y) H(1;z)
            - 6 H(0;z) H(1;y)
            - 8 H(0,0;y)
            - 8 H(0,0;z)\nonumber \\
&&
            + 2 H(0,1;y)
            - 6 H(0,1;z)
            - 8 H(0,1-z;y)
            - 6 H(1;y) H(1;z)
            - 4 H(1;z) H(1-z;y)\nonumber \\
&&
            - 6 H(1,0;y)
            - 6 H(1,0;z)
            - H(1,1;y)
            - 11 H(1,1;z)
            - 6 H(1,1-z;y)\nonumber \\
&&
            + 2 H(1-z,1;y)
            - 6 H(1-z,1-z;y)
                            \big] \ ,\\
f_{7.7,4}(y,z) &=& -\frac{5}{4}\ , \\
f_{7.7,3}(y,z) &=& 
          + \frac{5}{2} H(0;y)
          + \frac{3}{2} H(0;z)
          + \frac{1}{2} H(1;z)
          + \frac{1}{2} H(1-z;y) \ , \\
f_{7.7,2}(y;z) &=& 
          - 3 H(0;y) H(0;z)
          - H(0;y) H(1;z)
          + H(0;z) H(1-z;y)
          - 5 H(0,0;y)
          - 3 H(0,0;z)\nonumber \\
&&
          + 3 H(0,1;z)
          - H(0,1-z;y)
          + 4 H(1;z) H(1-z;y)
          + 2 H(1;z) H(z;y)
          - 3 H(1,0;y)\nonumber \\
&&
          + 4 H(1,1;z)
          - H(1-z,0;y)
          + 4 H(1-z,1-z;y)
          + 2 H(z,1-z;y)
          -\frac{\pi^2}{6}\ , \\
f_{7.7,1}(y;z) &=& 
          - 8 H(0;y) H(1,1;z)
          + 4 H(0;z) H(1,0;y)
          + 4 H(0;z) H(z,1-z;y)
          + 6 H(0,0;y) H(0;z)\nonumber \\
&&
          + 2 H(0,0;y) H(1;z)
          + 6 H(0,0;z) H(0;y)
          - 2 H(0,0;z) H(1-z;y)
          + 10 H(0,0,0;y)\nonumber \\
&&
          + 6 H(0,0,0;z)
          + 9 H(0,0,1;z)
          + 2 H(0,0,1-z;y)
          - 6 H(0,1;z) H(0;y)\nonumber \\
&&
          + 9 H(0,1;z) H(1-z;y)
          + 7 H(0,1;z) H(z;y)
          + 6 H(0,1,0;y)
          + 9 H(0,1,0;z)\nonumber \\
&&
          + 6 H(0,1,1;z)
          - 2 H(0,1-z;y) H(0;z)
          - 8 H(0,1-z;y) H(1;z)
          + 2 H(0,1-z,0;y)\nonumber \\
&&
          - 8 H(0,1-z,1-z;y)
          - 4 H(0,z;y) H(1;z)
          - 4 H(0,z,1-z;y)
          - 8 H(1;z) H(1-z,0;y)\nonumber \\
&&
          + 11 H(1;z) H(1-z,1-z;y)
          + 9 H(1;z) H(1-z,z;y)
          - 4 H(1;z) H(z,0;y)\nonumber \\
&&
          + 6 H(1;z) H(z,1-z;y)
          + 3 H(1;z) H(z,z;y)
          + 4 H(1,0;y) H(1;z)
          + 2 H(1,0;z) H(1-z;y)\nonumber \\
&&
          + 4 H(1,0;z) H(z;y)
          + 6 H(1,0,0;y)
          + 9 H(1,0,1;z)
          + 4 H(1,0,1-z;y)\nonumber \\
&&
          + 11 H(1,1;z) H(1-z;y)
          + 6 H(1,1;z) H(z;y)
          + H(1,1,0;y)
          + 11 H(1,1,1;z)\nonumber \\
&&
          + 4 H(1,1-z,0;y)
          + 2 H(1-z,0,0;y)
          - 8 H(1-z,0,1-z;y)
          + 6 H(1-z,1,0;y)\nonumber \\
&&
          - 8 H(1-z,1-z,0;y)
          + 11 H(1-z,1-z,1-z;y)
          + 9 H(1-z,z,1-z;y)\nonumber \\
&&
          - 4 H(z,0,1-z;y)
          - 4 H(z,1-z,0;y)
          + 6 H(z,1-z,1-z;y)
          + 3 H(z,z,1-z;y)\nonumber \\
&&
          + \frac{37}{2} \zeta_3 
          + \frac{\pi^2}{6} \big[
             + 2 H(0;y)
             + 3 H(0;z)
             + H(1;y)
             - 2 H(1;z)
                            \big] \ ,\\
f_{7.7,0}(y;z) &=& 
          - 18 H(0;y) H(1,0,1;z)
          - 22 H(0;y) H(1,1,1;z)
          - 8 H(0;z) H(1,0,0;y)\nonumber \\
&&
          - 2 H(0;z) H(1,0,1-z;y)
          - 8 H(0;z) H(1,1-z,0;y)
          - 2 H(0;z) H(1-z,0,1-z;y)\nonumber \\
&&
          - 2 H(0;z) H(1-z,1-z,1-z;y)
          + 8 H(0;z) H(z,0,1-z;y)
          - 2 H(0;z) H(z,1-z,0;y)\nonumber \\
&&
          + 4 H(0;z) H(z,1-z,1-z;y)
          - 12 H(0,0;y) H(0,0;z)
          + 12 H(0,0;y) H(0,1;z)\nonumber \\
&&
          + 16 H(0,0;y) H(1,1;z)
          + 4 H(0,0;z) H(0,1-z;y)
          - 8 H(0,0;z) H(1,0;y)\nonumber \\
&&
          - 8 H(0,0;z) H(z,1-z;y)
          - 12 H(0,0,0;y) H(0;z)
          - 4 H(0,0,0;y) H(1;z)\nonumber \\
&&
          - 12 H(0,0,0;z) H(0;y)
          + 4 H(0,0,0;z) H(1-z;y)
          - 20 H(0,0,0,0;y)\nonumber \\
&&
          - 12 H(0,0,0,0;z)
          + 21 H(0,0,0,1;z)
          - 4 H(0,0,0,1-z;y)
          - 18 H(0,0,1;z) H(0;y)\nonumber \\
&&
          - 12 H(0,0,1;z) H(1;y)
          + 13 H(0,0,1;z) H(1-z;y)
          + 5 H(0,0,1;z) H(z;y)\nonumber \\
&&
          - 12 H(0,0,1,0;y)
          + 18 H(0,0,1,1;z)
          + 4 H(0,0,1-z;y) H(0;z)\nonumber \\
&&
          + 16 H(0,0,1-z;y) H(1;z)
          - 4 H(0,0,1-z,0;y)
          + 16 H(0,0,1-z,1-z;y)\nonumber \\
&&
          + 8 H(0,0,z;y) H(1;z)
          + 8 H(0,0,z,1-z;y)
          - 18 H(0,1;z) H(0,1-z;y)\nonumber \\
&&
          - 14 H(0,1;z) H(0,z;y)
          + 10 H(0,1;z) H(1,0;y)
          - 12 H(0,1;z) H(1,1-z;y)\nonumber \\
&&
          - 18 H(0,1;z) H(1-z,0;y)
          + 21 H(0,1;z) H(1-z,1-z;y)
          - 3 H(0,1;z) H(1-z,z;y)\nonumber \\
&&
          - 8 H(0,1;z) H(z,0;y)
          + 4 H(0,1;z) H(z,1-z;y)
          - 3 H(0,1;z) H(z,z;y)\nonumber \\
&&
          - 8 H(0,1,0;y) H(0;z)
          - 8 H(0,1,0;y) H(1;z)
          - 18 H(0,1,0;z) H(0;y)\nonumber \\
&&
          - 6 H(0,1,0;z) H(1;y)
          + 2 H(0,1,0;z) H(1-z;y)
          + 2 H(0,1,0;z) H(z;y)\nonumber \\
&&
          - 12 H(0,1,0,0;y)
          - 18 H(0,1,0,0;z)
          - 8 H(0,1,0,1-z;y)
          - 12 H(0,1,1;z) H(0;y)\nonumber \\
&&
          + 19 H(0,1,1;z) H(1-z;y)
          + 6 H(0,1,1;z) H(z;y)
          - 2 H(0,1,1,0;y)
          - 21 H(0,1,1,0;z)\nonumber \\
&&
          + 12 H(0,1,1,1;z)
          - 8 H(0,1,1-z,0;y)
          - 4 H(0,1-z;y) H(1,0;z)\nonumber \\
&&
          - 22 H(0,1-z;y) H(1,1;z)
          + 16 H(0,1-z,0;y) H(1;z)
          - 4 H(0,1-z,0,0;y)\nonumber \\
&&
          + 16 H(0,1-z,0,1-z;y)
          - 12 H(0,1-z,1,0;y)
          - 22 H(0,1-z,1-z;y) H(1;z)\nonumber \\
&&
          + 16 H(0,1-z,1-z,0;y)
          - 22 H(0,1-z,1-z,1-z;y)
          - 18 H(0,1-z,z;y) H(1;z)\nonumber \\
&&
          - 18 H(0,1-z,z,1-z;y)
          - 8 H(0,z;y) H(1,0;z)
          - 12 H(0,z;y) H(1,1;z)\nonumber \\
&&
          + 8 H(0,z,0;y) H(1;z)
          + 8 H(0,z,0,1-z;y)
          - 8 H(0,z,1-z;y) H(0;z)\nonumber \\
&&
          - 12 H(0,z,1-z;y) H(1;z)
          + 8 H(0,z,1-z,0;y)
          - 12 H(0,z,1-z,1-z;y)\nonumber \\
&&
          - 6 H(0,z,z;y) H(1;z)
          - 6 H(0,z,z,1-z;y)
          + 16 H(1;z) H(1-z,0,0;y)\nonumber \\
&&
          - 22 H(1;z) H(1-z,0,1-z;y)
          - 16 H(1;z) H(1-z,0,z;y)\nonumber \\
&&
          - 22 H(1;z) H(1-z,1-z,0;y)
          + 25 H(1;z) H(1-z,1-z,1-z;y)\nonumber \\
&&
          + 23 H(1;z) H(1-z,1-z,z;y)
          + 21 H(1;z) H(1-z,z,1-z;y)\nonumber \\
&&
          - 3 H(1;z) H(1-z,z,z;y)
          + 8 H(1;z) H(z,0,0;y)
          - 4 H(1;z) H(z,0,1-z;y)\nonumber \\
&&
          - 16 H(1;z) H(z,0,z;y)
          - 4 H(1;z) H(z,1-z,0;y)
          + 14 H(1;z) H(z,1-z,1-z;y)\nonumber \\
&&
          + 2 H(1;z) H(z,z,1-z;y)
          - 3 H(1;z) H(z,z,z;y)
          - 8 H(1,0;y) H(1,0;z)\nonumber \\
&&
          + 10 H(1,0;y) H(1,1;z)
          - 6 H(1,0;z) H(1,1-z;y)
          - 2 H(1,0;z) H(z,0;y)\nonumber \\
&&
          - 6 H(1,0;z) H(z,1-z;y)
          - 8 H(1,0,0;y) H(1;z)
          - 4 H(1,0,0;z) H(1-z;y)\nonumber \\
&&
          - 8 H(1,0,0;z) H(z;y)
          - 12 H(1,0,0,0;y)
          + 15 H(1,0,0,1;z)\nonumber \\
&&
          - 8 H(1,0,0,1-z;y)
          - 12 H(1,0,1;z) H(1;y)
          + 21 H(1,0,1;z) H(1-z;y)\nonumber \\
&&
          + 4 H(1,0,1;z) H(z;y)
          + 2 H(1,0,1,0;y)
          + 21 H(1,0,1,1;z)\nonumber \\
&&
          + 10 H(1,0,1-z;y) H(1;z)
          - 8 H(1,0,1-z,0;y)
          + 10 H(1,0,1-z,1-z;y)\nonumber \\
&&
          + 12 H(1,0,z;y) H(1;z)
          + 12 H(1,0,z,1-z;y)
          - 22 H(1,1;z) H(1-z,0;y)\nonumber \\
&&
          + 25 H(1,1;z) H(1-z,1-z;y)
          + 21 H(1,1;z) H(1-z,z;y)
          - 4 H(1,1;z) H(z,0;y)\nonumber \\
&&
          + 14 H(1,1;z) H(z,1-z;y)
          + 2 H(1,1;z) H(z,z;y)
          - 6 H(1,1,0;z) H(1;y)\nonumber \\
&&
          + 4 H(1,1,0;z) H(1-z;y)
          - 6 H(1,1,0;z) H(z;y)
          - 2 H(1,1,0,0;y)\nonumber \\
&&
          + 21 H(1,1,0,1;z)
          + 25 H(1,1,1;z) H(1-z;y)
          + 14 H(1,1,1;z) H(z;y)\nonumber \\
&&
          - H(1,1,1,0;y)
          + 25 H(1,1,1,1;z)
          + 10 H(1,1-z,0;y) H(1;z)\nonumber \\
&&
          - 8 H(1,1-z,0,0;y)
          + 10 H(1,1-z,0,1-z;y)
          - 18 H(1,1-z,1,0;y)\nonumber \\
&&
          + 10 H(1,1-z,1-z,0;y)
          - 12 H(1,1-z,z;y) H(1;z)
          - 12 H(1,1-z,z,1-z;y)\nonumber \\
&&
          - 4 H(1-z,0,0,0;y)
          + 16 H(1-z,0,0,1-z;y)
          - 12 H(1-z,0,1,0;y)\nonumber \\
&&
          + 16 H(1-z,0,1-z,0;y)
          - 22 H(1-z,0,1-z,1-z;y)
          - 16 H(1-z,0,z,1-z;y)\nonumber \\
&&
          - 12 H(1-z,1,0,0;y)
          + 6 H(1-z,1,1,0;y)
          + 16 H(1-z,1-z,0,0;y)\nonumber \\
&&
          - 22 H(1-z,1-z,0,1-z;y)
          + 18 H(1-z,1-z,1,0;y)
          - 22 H(1-z,1-z,1-z,0;y)\nonumber \\
&&
          + 25 H(1-z,1-z,1-z,1-z;y)
          + 23 H(1-z,1-z,z,1-z;y)\nonumber \\
&&
          + 21 H(1-z,z,1-z,1-z;y)
          - 3 H(1-z,z,z,1-z;y)
          + 8 H(z,0,0,1-z;y)\nonumber \\
&&
          - 6 H(z,0,1,0;y)
          + 8 H(z,0,1-z,0;y)
          - 4 H(z,0,1-z,1-z;y)\nonumber \\
&&
          - 16 H(z,0,z,1-z;y)
          + 8 H(z,1-z,0,0;y)
          - 4 H(z,1-z,0,1-z;y)\nonumber \\
&&
          + 6 H(z,1-z,1,0;y)
          - 4 H(z,1-z,1-z,0;y)
          + 14 H(z,1-z,1-z,1-z;y)\nonumber \\
&&
          + 2 H(z,z,1-z,1-z;y)
          - 3 H(z,z,z,1-z;y)
          + \frac{\mbox{31}\pi^4}{60}\nonumber \\
&&
          + \zeta_3 \big[
             - 37 H(0;y)
             - 24 H(0;z)
             - 11 H(1;y)
             - 17 H(1;z)
             - 9 H(1-z;y)
                    \big]\nonumber \\
&&
          +\frac{\pi^2}{6} \big[
             - 6 H(0;y) H(0;z)
             + 4 H(0;y) H(1;z)
             - 6 H(0;z) H(1;y)
             + 2 H(0;z) H(1-z;y)\nonumber \\
&&
             - 4 H(0,0;y)
             - 6 H(0,0;z)
             - 2 H(0,1;y)
             - 15 H(0,1;z)
             - 6 H(1;y) H(1;z)\nonumber \\
&&
             - 2 H(1;z) H(z;y)
             - 6 H(1,0;y)
             - H(1,1;y)
             - 4 H(1,1;z)
             - 6 H(1,1-z;y)\nonumber \\
&&
             + 4 H(1-z,0;y)
             + 6 H(1-z,1;y)
             - 4 H(1-z,1-z;y)
             - 2 H(z,1-z;y)
                           \big]\ , \\
f_{7.8,4}(y,z) &=& 0 \ ,\\
f_{7.8,3}(y,z) &=& 0 \ , \\
f_{7.8,2}(y,z) &=& 0 \ , \\
f_{7.8,1}(y,z) &=&           
          - 2 H(0;y) H(1,1;z)
          + 2 H(0;z) H(1-z,1-z;y)
          + 2 H(0,1;z) H(1-z;y)
          + 2 H(0,1,1;z)\nonumber \\
&&
          - 2 H(0,1-z;y) H(1;z)
          - 2 H(0,1-z,1-z;y)
          - 2 H(1;z) H(1-z,0;y)\nonumber \\
&&
          + 2 H(1,0;z) H(1-z;y)
          + 2 H(1,0,1;z)
          + 2 H(1,1,0;z)
          - 2 H(1-z,0,1-z;y)\nonumber \\
&&
          - 2 H(1-z,1-z,0;y)\ ,  \\
f_{7.8,0}(y;z) &=&           
          - 8 H(0;y) H(1,0,1;z)
          - 4 H(0;y) H(1,1,0;z)
          - 6 H(0;y) H(1,1,1;z)\nonumber \\
&&
          + 4 H(0;z) H(1-z,1,0;y)
          + 6 H(0;z) H(1-z,1-z,1-z;y)\nonumber \\
&&
          + 8 H(0;z) H(1-z,z,1-z;y)
          - 4 H(0;z) H(z,1-z,1-z;y)
          + 4 H(0,0;y) H(1,1;z)\nonumber \\
&&
          - 4 H(0,0;z) H(1-z,1-z;y)
          + 12 H(0,0,1;z) H(1-z;y)
          - 4 H(0,0,1,1;z)\nonumber \\
&&
          + 4 H(0,0,1-z;y) H(1;z)
          + 4 H(0,0,1-z,1-z;y)
          - 4 H(0,1;z) H(0,1-z;y)\nonumber \\
&&
          - 8 H(0,1;z) H(1-z,0;y)
          + 6 H(0,1;z) H(1-z,1-z;y)
          + 8 H(0,1;z) H(1-z,z;y)\nonumber \\
&&
          - 4 H(0,1;z) H(z,1-z;y)
          - 4 H(0,1,0;y) H(0;z)
          - 4 H(0,1,0;y) H(1;z)\nonumber \\
&&
          + 4 H(0,1,0;z) H(1-z;y)
          + 4 H(0,1,0,1;z)
          - 4 H(0,1,0,1-z;y)\nonumber \\
&&
          + 6 H(0,1,1;z) H(1-z;y)
          - 4 H(0,1,1;z) H(z;y)
          - 4 H(0,1,1,0;y)
          + 6 H(0,1,1,1;z)\nonumber \\
&&
          - 4 H(0,1,1-z,0;y)
          + 4 H(0,1-z;y) H(1,0;z)
          - 6 H(0,1-z;y) H(1,1;z)\nonumber \\
&&
          + 4 H(0,1-z,0;y) H(0;z)
          + 4 H(0,1-z,0;y) H(1;z)
          + 4 H(0,1-z,0,1-z;y)\nonumber \\
&&
          + 4 H(0,1-z,1-z;y) H(0;z)
          - 6 H(0,1-z,1-z;y) H(1;z)
          + 4 H(0,1-z,1-z,0;y)\nonumber \\
&&
          - 6 H(0,1-z,1-z,1-z;y)
          - 8 H(0,1-z,z;y) H(1;z)
          - 8 H(0,1-z,z,1-z;y)\nonumber \\
&&
          - 4 H(0,z;y) H(1,1;z)
          - 4 H(0,z,1-z;y) H(1;z)
          - 4 H(0,z,1-z,1-z;y)\nonumber \\
&&
          + 4 H(1;z) H(1-z,0,0;y)
          - 6 H(1;z) H(1-z,0,1-z;y)
          - 8 H(1;z) H(1-z,0,z;y)\nonumber \\
&&
          + 4 H(1;z) H(1-z,1,0;y)
          - 6 H(1;z) H(1-z,1-z,0;y)
          - 8 H(1;z) H(1-z,z,0;y)\nonumber \\
&&
          + 4 H(1;z) H(z,0,1-z;y)
          + 4 H(1;z) H(z,1-z,0;y)
          - 4 H(1,0;z) H(1-z,0;y)\nonumber \\
&&
          + 2 H(1,0;z) H(1-z,1-z;y)
          + 8 H(1,0;z) H(1-z,z;y)
          - 4 H(1,0;z) H(z,1-z;y)\nonumber \\
&&
          - 4 H(1,0,0;z) H(1-z;y)
          + 12 H(1,0,0,1;z)
          + 2 H(1,0,1;z) H(1-z;y)\nonumber \\
&&
          - 4 H(1,0,1;z) H(z;y)
          + 4 H(1,0,1,0;z)
          + 6 H(1,0,1,1;z)
          - 6 H(1,1;z) H(1-z,0;y)\nonumber \\
&&
          + 4 H(1,1;z) H(z,0;y)
          - 6 H(1,1,0;z) H(1-z;y)
          - 4 H(1,1,0;z) H(z;y)
          - 4 H(1,1,0,0;z)\nonumber \\
&&
          + 2 H(1,1,0,1;z)
          - 6 H(1,1,1,0;z)
          + 4 H(1-z,0,0,1-z;y)
          + 4 H(1-z,0,1-z,0;y)\nonumber \\
&&
          - 6 H(1-z,0,1-z,1-z;y)
          - 8 H(1-z,0,z,1-z;y)
          + 4 H(1-z,1,0,1-z;y)\nonumber \\
&&
          + 4 H(1-z,1,1,0;y)
          + 4 H(1-z,1,1-z,0;y)
          + 4 H(1-z,1-z,0,0;y)\nonumber \\
&&
          - 6 H(1-z,1-z,0,1-z;y)
          + 4 H(1-z,1-z,1,0;y)
          - 6 H(1-z,1-z,1-z,0;y)\nonumber \\
&&
          - 8 H(1-z,z,0,1-z;y)
          - 8 H(1-z,z,1-z,0;y)
          + 4 H(z,0,1-z,1-z;y)\nonumber \\
&&
          + 4 H(z,1-z,0,1-z;y)
          + 4 H(z,1-z,1-z,0;y) 
          +\frac{\pi^2}{6} \big[
          - 4 H(0,1;y)
          + 4 H(0,1-z;y)\nonumber \\
&&
          - 4 H(1;z) H(1-z;y)
          - 4 H(1,1;z)
          + 4 H(1-z,1;y)
                           \big]   \ .
\end{eqnarray}

Both seven-propagator master
integrals without scalar products in the numerator 
(\ref{eq:npunit1}), (\ref{eq:npunit2}) 
were first considered by Binoth and Heinrich~\cite{num}. Using a dedicated 
algebraic procedure to separate the structure of infrared divergencies, 
the $1/\e^4$ and $1/\e^3$ terms of these integrals were obtained 
analytically in~\cite{num}, the remaining terms were computed numerically 
order by order in $\e$. We agree with the
results for the $1/\e^4$ and $1/\e^3$ terms of~\cite{num}. 

More recently, Smirnov has considered~\cite{smirnp} the problem of evaluating 
 one of the  seven-propagator master
integrals (\ref{eq:npunit1}). The result, quoted in~\cite{smirnp}, contains 
explicit expressions for all divergent parts of (\ref{eq:npunit1}), which we 
confirm. The finite part of (\ref{eq:npunit1}), however, is expressed as a 
three-dimensional Mellin--Barnes integral, which we are unable to transform 
into two-dimensional harmonic polylogarithms. A direct comparison 
with the finite part of~\cite{smirnp} is therefore not possible. 
Based on general considerations about the analytic 
structure of the results obtained in~\cite{smirnp}, Smirnov moreover argues
that two-dimensional harmonic polylogarithms could be insufficient 
to describe the finite part  of (\ref{eq:npunit1}) and calls for the 
introduction of three-dimensional harmonic polylogarithms -- but that is 
not the case, as shown by our results. 

\section{Conclusion}
\label{sec:conc}
\setcounter{equation}{0}

Two-loop four-point functions with one off-shell leg are an important 
ingredient for the calculation of next-to-next-to-leading order 
corrections to three-jet production and related observables 
in electron--positron annihilation. By exploiting 
integration-by-parts~\cite{hv,chet1,chet2} and Lorentz invariance~\cite{gr} 
identities, one can express the loop integrals appearing in these 
functions as a linear combination of a small set of master 
integrals, which are scalar 
functions of the external invariants. Their determination was 
up to now a major obstacle to further progress in
next-to-next-to-leading order calculations. 

These master integrals fulfil inhomogeneous differential equations~\cite{gr} 
in their external invariants, which can be used for their computation. 
In this approach, one solves the differential equations by starting from 
an appropriate ansatz, which is determined from the homogeneous part 
of the equation. Once this ansatz is found, the equations are expanded in 
$\e = (4-d)/2$ and then solved by repeated quadratures, 
yielding a general solution that still contains free 
constants of integration. These 
constants are determined from boundary conditions 
to the same differential equations. 
Using this method, we have computed 
the master integrals for planar two-loop four-point functions 
with one off-shell leg and presented it
in a previous publication~\cite{grplanar}. In the 
present work, we have 
now derived the corresponding non-planar master integrals as well. 
The results are expressed as a Laurent series in $\e$, with coefficients 
containing 2dHPL, which were introduced 
in~\cite{grplanar} as generalization of harmonic polylogarithms~\cite{hpl} 
and of 
Nielsen's generalized 
polylogarithms~\cite{nielsen}. All 2dHPL up to weight 3, as those 
appearing in the divergent parts of the master integrals, can be expressed 
in terms of generalized  polylogarithms of complicated arguments, 
see~\cite{grplanar}, while the 2dHPL of weight 4 appearing in the finite 
parts of the master integrals involve 
a one-dimensional integral over a combination of generalized 
polylogarithms of weight 3. 

The results presented in this paper 
 complete the calculation of master integrals needed 
for the computation of two-loop four-point functions with one off-shell 
leg. These functions are a crucial ingredient for
 the virtual next-to-next-to-leading order corrections to processes 
such as three-jet production in electron-positron annihilation, 
two-plus-one-jet production in deep inelastic scattering, and 
vector-boson-plus-jet production at hadron colliders. The results given 
in~\cite{grplanar} and in the present work correspond to the 
kinematical situation of  three-jet production in electron--positron 
annihilation, where all master integrals are real (up 
to a complex normalization factor). 
Using the prescription for the analytic continuation of 
two-dimensional harmonic polylogarithms derived in the 
appendix of~\cite{grplanar}, our results can be continued into
any other kinematic region of interest. 

Equipped with the complete  set of master integrals, it should soon become 
possible to compute the two-loop virtual corrections to the 
above-mentioned processes. However, it must be kept in mind that these 
corrections form only part of a full next-to-next-to-leading order 
calculation, which also has to include the one-loop corrections 
to processes with one soft or collinear real parton~\cite{onel1,onel2,onel3},
 as well as 
tree-level processes with two soft or collinear partons~\cite{gg}. 
Only after summing all these contributions (and including 
terms from the renormalization of parton distributions 
for processes with partons in the initial state), do
the divergent terms cancel 
among one another. The remaining finite terms have to be combined 
into a numerical programme implementing the 
experimental definition of jet observables and event-shape variables. 
A first calculation involving the above features 
was presented for the case of 
photon-plus-one-jet final 
states in electron--positron annihilation in~\cite{gg},
thus demonstrating the feasibility of this type of calculations.

{\bf Note added:} After this paper was submitted for publication, we 
have carried out a detailed comparison with the purely numerical 
results of~\cite{num}. We find full agreement with the values quoted 
in~\cite{num}
for the master integrals (3.21) and (3.38) at specific points of the 
kinematical invariants, as well as with values for the master integral
(3.9) provided by the authors of~\cite{num}. We would like to thank 
Thomas Binoth and Gudrun Heinrich for their assistance with this
comparison.

\section*{Acknowledgement} 
We are grateful to Jos Vermaseren for his assistance in the use of 
the algebraic program FORM~\cite{form}, which was intensively used in 
all the steps of the calculation, as well as for useful communication 
concerning harmonic sums and harmonic polylogarithms.


\begin{thebibliography}{99}
\bibitem{dreg1}
C.G.\ Bollini and J.J.\ Giambiagi, Nuovo Cim.\ {\bf 12B} (1972) 20.

\bibitem{dreg2}
G.M.\ Cicuta and E.\ Montaldi, Nuovo Cim.\ Lett.\ {\bf 4} (1972) 329.

\bibitem{hv}
G.\ 't Hooft and M.\ Veltman, Nucl.\ Phys.\ {\bf B44} (1972) 189.

\bibitem{chet1}
F.V.\ Tkachov, Phys.\ Lett.\ {\bf 100B} (1981) 65.
\bibitem{chet2}
K.G.\ Chetyrkin and F.V.\ Tkachov, Nucl.\ Phys.\ {\bf B192} (1981) 159.

\bibitem{gr} 
T.\ Gehrmann and E.\ Remiddi, Nucl.\ Phys.\
{\bf B580} (2000) 485.

\bibitem{onshell1}
V.A.\ Smirnov, Phys.\ Lett.\ {\bf B460} (1999) 397.
\bibitem{onshell2}
J.B.\ Tausk, Phys.\ Lett.\ {\bf B469} (1999) 225.
\bibitem{onshell3}
V.A.\ Smirnov and O.L.\ Veretin,  Nucl.\ Phys.\ {\bf B566}
(2000) 469.
\bibitem{onshell4}
C.\ Anastasiou, T.\ Gehrmann, C.\ Oleari, E.\ Remiddi and 
J.B.\ Tausk, Nucl.\ Phys.\
{\bf B580} (2000) 577.
\bibitem{onshell5}
T.~Gehrmann and E.~Remiddi, Nucl.\ Phys.\ {\bf B} (Proc.\ Suppl.)
{\bf 89} (2000) 251.
\bibitem{onshell6}
C.\ Anastasiou, J.B.\ Tausk and M.E.\ Tejeda-Yeomans, 
Nucl.\ Phys.\ {\bf B} (Proc.\ Suppl.) {\bf 89} (2000) 262.

\bibitem{m1}
Z.~Bern, L.~Dixon and A.~Ghinculov, Phys.\ Rev.\ {\bf D63} (2001) 053007.

\bibitem{m2}
C.\ Anastasiou, E.W.N.~Glover, C.\ Oleari and M.E.\ Tejeda-Yeomans,
Nucl.\ Phys.\ {\bf B601} (2001) 318; {\bf B601} (2001) 347; Phys.\ Lett.\
{\bf B506} (2001) 59.


\bibitem{smirnew}
V.A.~Smirnov, Phys.\ Lett.\ {\bf B491} (2000) 130.

\bibitem{grplanar}
T.\ Gehrmann and E.\ Remiddi, Nucl.~Phys.~{\bf B601} (2001) 248.

\bibitem{smirnp}
V.A.\ Smirnov, Phys.\ Lett.\ {\bf B500} (2001) 330.

\bibitem{num}
T.\ Binoth and G.\ Heinrich, Nucl.\ Phys.\ {\bf B585} (2000) 741.

\bibitem{hpl}
E.\ Remiddi and J.A.M.\ Vermaseren, Int.\ J.\ Mod.\ Phys.\ {\bf A15}
(2000) 725.

\bibitem{nielsen}
N.~Nielsen, Nova Acta Leopoldiana (Halle) {\bf 90} (1909) 123.

\bibitem{bit}
K.S.\ K\"olbig, J.A.\ Mignaco and E.\ Remiddi, BIT {\bf 10} (1970) 38.


\bibitem{kotikov}
A.V.\ Kotikov, Phys.\ Lett.\ {\bf B254} (1991) 158.

\bibitem{remiddi}
E.\ Remiddi, Nuovo Cim.~{\bf 110A} (1997) 1435 (hep-th/9711188).

\bibitem{appl}
M.~Caffo, H.~Czy\.{z} and E.~Remiddi, Nuovo Cim.~{\bf 111A} (1998) 365
(hep-th/9805118), Nucl.\ Phys.\ {\bf B581} (2000) 274.

\bibitem{form}
J.A.M.~Vermaseren, {\it Symbolic Manipulation with FORM}, Version 2,
CAN, Amsterdam, 1991.
%
\bibitem{maple}
{\it MAPLE V Release 3},  Copyright 1981-1994 
by Waterloo Maple Software and the University of Waterloo.
%

\bibitem{kl1}
R.J.\ Gonsalves, Phys.\ Rev.\ {\bf D28} (1983) 1542.
\bibitem{kl2}
G.\ Kramer and B.\ Lampe, J.\ Math.\ Phys.\ {\bf 28} (1987) 945.

\bibitem{catani}
S.\ Catani, Phys.\ Lett.\ {\bf B427} (1998) 161.

\bibitem{onel1}
D.A.~Kosower and P.~Uwer, Nucl.\ Phys.\ {\bf B563} (1999) 477.

\bibitem{onel2}
Z.\ Bern, V.\ Del Duca, W.B.\ Kilgore and C.R.\ Schmidt, Phys.\ Rev.\
{\bf D60} (1999) 116001.

\bibitem{onel3}
S.\ Catani and M.\ Grazzini, Nucl.\ Phys.\  {\bf B591} (2000) 435.

\bibitem{gg}
A.~Gehrmann-De Ridder and E.W.N.~Glover, Nucl.~Phys.\ {\bf B517} (1998) 269.

\end{thebibliography}
\end{document}